\documentclass[10pt,conference]{IEEEtran}

\usepackage{cite}
\usepackage{amsmath,amssymb,amsfonts}
\usepackage{algorithmic}
\usepackage{graphicx}
\usepackage{textcomp}
\usepackage{xcolor}
\usepackage{subcaption}
\usepackage{siunitx}
\usepackage{hyperref}
\hypersetup{
  pdftitle={The Impact of Qubit Connectivity on Quantum Advantage in Noisy IQP Circuits},
  pdfauthor={Leonardo Placidi, Enrico Rinaldi, Keisuke Fujii, Chen-Yu Liu},
  pdfsubject={Connectivity and classical simulatability of noisy IQP circuits},
  pdfkeywords={IQP circuits, qubit connectivity, quantum advantage, classical simulatability}
}
\usepackage{booktabs}
\usepackage{verbatim} 
\usepackage{subcaption}

\hypersetup{
  colorlinks   = true, 
  urlcolor     = blue, 
  linkcolor    = blue, 
  citecolor   = blue 
}

\def\BibTeX{{\rm B\kern-.05em{\sc i\kern-.025em b}\kern-.08em
    T\kern-.1667em\lower.7ex\hbox{E}\kern-.125emX}}
\begin{document}

\title{The Impact of Qubit Connectivity on Quantum Advantage in Noisy IQP Circuits
}
\author{
\IEEEauthorblockN{
    Leonardo Placidi\IEEEauthorrefmark{1}\IEEEauthorrefmark{3}, 
    Enrico Rinaldi\IEEEauthorrefmark{2},
    Keisuke Fujii\IEEEauthorrefmark{3},
    Chen-Yu Liu\IEEEauthorrefmark{2}
}
\IEEEauthorblockA{\IEEEauthorrefmark{1} Quantinuum, Otemachi Financial City Grand Cube, Tokyo, Japan}
\IEEEauthorblockA{\IEEEauthorrefmark{2} Quantinuum, Partnership House, London, UK}
\IEEEauthorblockA{\IEEEauthorrefmark{3} The University of Osaka, Osaka, Japan}
\IEEEauthorblockA{
\texttt{
leonardo.placidi@quantinuum.com, 
enrico.rinaldi@quantinuum.com,}}
\IEEEauthorblockA{
\texttt{fujii.keisuke.es@osaka-u.ac.jp,
chen-yu.liu@quantinuum.com
}
}
}

\maketitle
\bstctlcite{IEEEexample:BSTcontrol}

\begin{abstract}
Instantaneous Quantum Polynomial-time (IQP) circuits are a candidate for demonstrating near-term quantum advantage, as their sampling task is believed to be classically hard in the ideal theoretical setting under standard complexity-theoretic assumptions. In noisy implementations, however, this hardness can disappear once circuit depth exceeds a noise-dependent critical threshold. We show that qubit connectivity is a key parameter in this transition, since sparse architectures require additional routing to implement long-range interactions, thereby increasing compiled circuit depth. To make this explicit, we present a connectivity-aware analysis of compiled IQP circuits. For a fixed abstract IQP instance, different hardware connectivity graphs yield different compiled depths and thus different effective positions relative to the noisy-IQP simulatability boundary. We quantify this architecture-dependent shift using the compiled depth overhead and the corresponding simulatability margin. We combine analytic depth estimates for sparse geometries, including the two-dimensional grid, with native-gateset-aware compilation experiments across seven hardware-grounded experimental device models derived from publicly available topologies. To compare these device models under a unified empirical framework, we approximate the effective noise level primarily through reported two-qubit gate error rates. This lets us compare how much effective noise sparse and fully connected architectures can tolerate for the same position relative to the noisy-IQP simulatability boundary. Our results show that sparse connectivity requires a lower effective noise level to sustain the same margin relative to the noisy-IQP simulatability boundary, and they provide a quantitative framework for determining when compiled IQP experiments are likely to remain outside, or instead enter, the classically simulatable regime.
\end{abstract}


\section{Introduction}
Demonstrating quantum advantage in the near term remains challenging because candidate circuits must satisfy two requirements: they should be believed to be classically hard to simulate in the ideal theoretical setting, and they must also remain outside the classically simulatable regime after the effects of noise, limited connectivity, and compilation overhead are taken into account~\cite{Preskill2018quantumcomputingin}. One of the most studied proposals is Instantaneous Quantum Polynomial-time (IQP) circuits, whose sampling task is believed to be classically intractable in the ideal theoretical setting under standard complexity-theoretic assumptions (for example, the polynomial-hierarchy-collapse consequences if an efficient classical simulation of IQP were possible)\cite{bremner_classical_2010, Bremner_2016}. 
In the idealized setting, IQP circuit sampling plays a central role in theoretical proposals for near-term quantum advantage~\cite{liu2026generativequantumutilitycorrelationcomplexity, shen2026characterizingtrainabilityinstantaneousquantum}.

\begin{figure}[htbp]
    \centering
    \includegraphics[width=.49\textwidth]{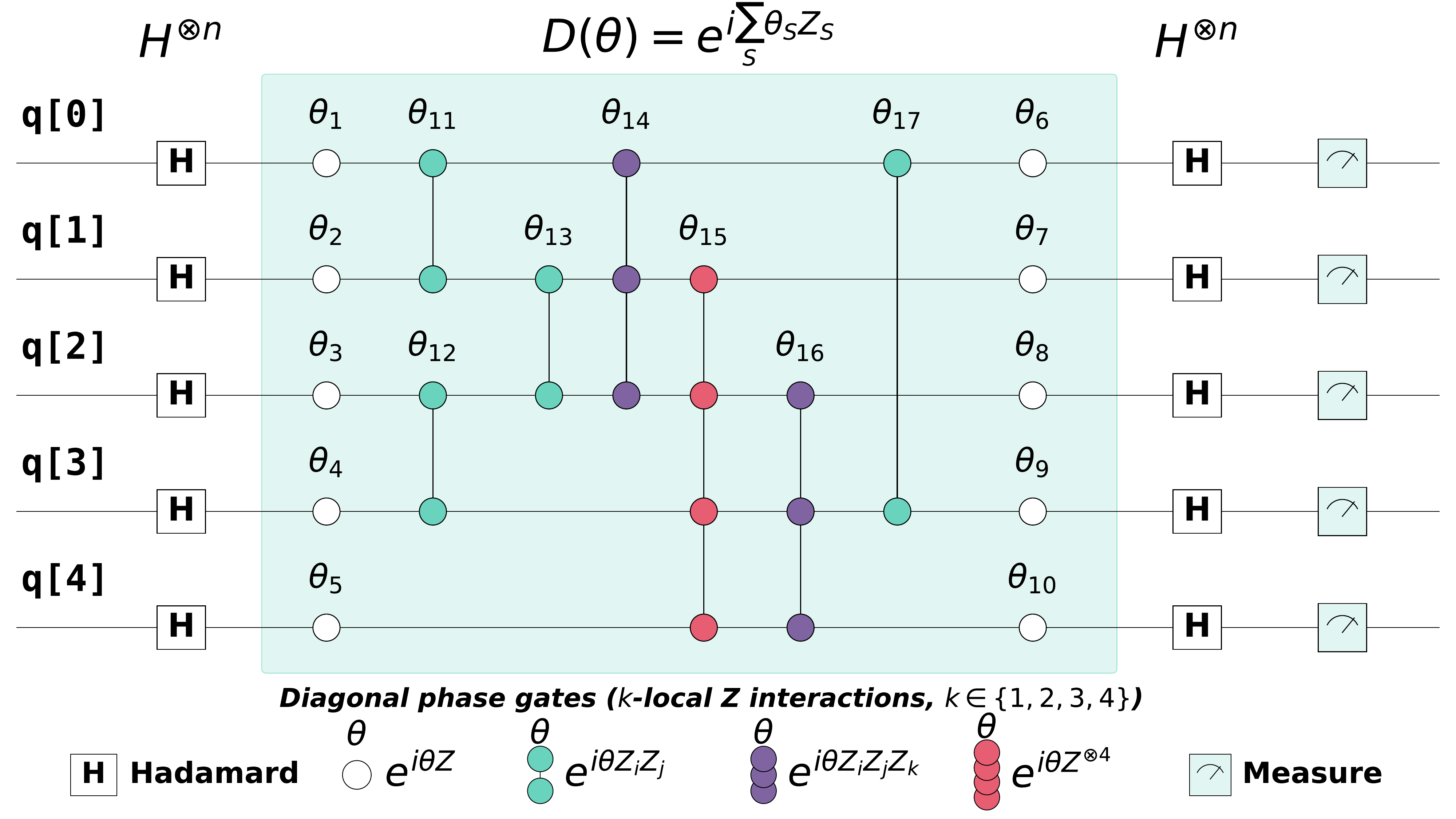}
    \caption{Structure of an IQP (Instantaneous Quantum Polynomial) circuit. 
    The circuit applies Hadamards to create superposition, then a diagonal unitary 
    $D(\theta) = e^{i\sum_S \theta_S Z_S}$ with $k$-local interactions (single-qubit $R_Z$, 
    two-qubit ZZ, three-qubit ZZZ, etc.), followed by Hadamards and measurement. 
    The diagonal gates commute and can be applied in any order, but compilation to sparse hardware may require connectivity-aware synthesis, routing, or additional native entangling operations.}
    \label{fig:iqp_circuit}
\end{figure}

Despite these ideal-case hardness results, a growing body of theory shows that sufficiently strong hardware noise can destroy the complexity-theoretic evidence for hardness and render IQP circuits efficiently classically simulatable in relevant noisy models~\cite{Fujii2016NoisyCommuting, Bremner_2017, rajakumar2025polynomial}. In particular, recent work identifies an efficiently classically simulatable regime, or ‘easy phase,’ for noisy IQP circuits\cite{rajakumar2025polynomial}: when physical noise is interleaved throughout the circuit at sufficient strength, or when circuit depth is sufficiently large relative to the noise level, the resulting output distributions admit polynomial-time classical simulation. Intuitively, noise suppresses quantum correlations and fragments the effective interaction structure of the circuit, so that the computation reduces to small connected components that can be handled classically. This reframes the central question from ``Is sampling from the output distribution of an IQP circuit classically hard?'' to ``Under what hardware-level conditions does an IQP implementation remain in a regime with complexity-theoretic evidence for hardness?''

Most discussions of noisy-to-easy transitions treat ``noise strength'' as an intrinsic device property, e.g., a per-gate Pauli error rate or a per-circuit-depth dephasing rate, and analyze hardness as a function of that parameter and circuit depth. But hardware connectivity introduces an additional, often dominant, effect: the same abstract IQP circuit does not compile to the same physical circuit across architectures.
This dependence of implemented circuits on device topology is already reflected in benchmarking approaches such as quantum volume, where connectivity and compilation directly influence the largest circuit size that can be successfully realized~\cite{Cross_2019}. In the context of IQP sampling, however, the implications of this architectural dependence for noisy-to-easy transitions have not been explicitly analyzed.

Prior work established complementary hardness and simulatability regimes for
noisy commuting circuits. Fujii and Tamate studied a decoherence-induced
boundary; Bremner \emph{et al.} identified sparse hard families and efficient
simulation under constant output noise; Rajakumar \emph{et al.} derived a
noise-dependent constant-depth simulation threshold; and Shravan
\emph{et al.} extended simulation to amplitude-damping noise
~\cite{Fujii2016NoisyCommuting,Bremner_2017,rajakumar2025polynomial,
shravan2026efficientsimulationnoisyiqp}. Broader volumetric and full-stack
benchmarks study width, depth and compilation tradeoffs
~\cite{Blume_Kohout_2020,Mills_2021,
tomesh2022supermarqscalablequantumbenchmark}, whereas we relate
hardware-dependent compiled depth directly to an explicit IQP
simulatability condition.

In connectivity-limited platforms (e.g.\ planar superconducting layouts), implementing nonlocal two-qubit interactions requires routing, typically realized through SWAP networks and additional entangling operations \cite{cowtan2019qubit}. Even when the abstract circuit is shallow and structured, routing can increase the two-qubit gate count and circuit depth. 
In contrast, architectures with near all-to-all or dynamically reconfigurable long-range connectivity, most notably trapped-ion and neutral-atom platforms, can often implement the same interaction pattern with substantially less routing overhead than fixed nearest-neighbor layouts \cite{decross2025computational, evered2023high}. 
Consequently, compilation overhead induced by limited connectivity can move an implementation closer to regimes where noisy IQP circuits become efficiently classically simulatable.

SWAP routing is not unique: connectivity-aware parity synthesis and Parity
Twine can reduce routing overhead~\cite{connectivity-aware,swapless}. Our experiments use SWAP-based compilation; the same margin analysis can be
reapplied to any synthesis once its compiled depth and a compatible
effective-noise proxy are specified.

This paper develops a connectivity-aware framework for analyzing noisy IQP implementations under a proxy effective-noise model and a known noisy-IQP simulatability criterion. Rather than treating noise as a single scalar independent of hardware architecture, we connect architecture to effective noise and compiled depth through compilation. The objective is not to re-prove IQP hardness or introduce a new advantage model, but to formulate precisely and evaluate quantitatively the claim that \textit{connectivity is a critical resource for keeping noisy IQP implementations outside the known efficiently simulatable regime.}
We frame this as a shift in the simulatability phase diagram: for a fixed abstract IQP circuit family, different hardware graphs yield different compiled depths and different effective noise parameters, thereby moving the physical experiment along the axis controlling known simulatability thresholds. 

Concretely, for each hardware graph, we study the compiled depth, routing-induced overhead, and resulting simulatability margin, which measures the signed distance between the compiled implementation and the noisy-IQP critical-depth boundary. This makes explicit how hardware-dependent compilation can shift a fixed abstract IQP instance toward the efficiently simulatable regime.

We summarize the paper's main contributions as follows:
\begin{enumerate}
    \item \textbf{Connectivity-aware framework for simulatability shifts in noisy IQP circuits.} We identify qubit connectivity and routing overhead as key variables in the noisy-to-easy transition for compiled IQP implementations.
    \item \textbf{Quantitative theory for simulatability shifts.} We relate hardware graph, compiled depth, and effective noise to the noisy-IQP phase boundary, and derive connectivity-dependent depth and noise-budget relations across architectures.
    \item \textbf{Native-gateset-aware study on seven hardware-grounded experimental models.} Using models derived from publicly available topologies and error-rate proxies, we show empirically how sparse connectivity increases routing overhead, inflates depth, and pushes implementations toward the classically simulatable regime.
\end{enumerate}

While compilation overhead is widely recognized as a practical concern of quantum utility, we argue it can be computationally decisive for assessing whether a noisy IQP experiment operates in the hard regime or the simulatable regime.

\section{Polynomial-Time Simulation of Noisy IQP Circuits}
\label{sec:ptsiqp}
We summarize the key theoretical mechanism underlying the ``easy phase'' of IQP sampling under physically motivated noise, following the recent simulatability result in \cite{rajakumar2025polynomial}. 
An $n$-qubit $k$-local IQP circuit can be written as the commuting unitary
\begin{equation}
U_{\mathrm{IQP}}
= \exp\!\Bigl(i\!\!\sum_{\substack{s\subseteq [n]\\1\le |s|\le k}} \theta_s\, X_s\Bigr)
= H^{\otimes n}\,
\exp\!\Bigl(i\!\!\sum_{\substack{s\subseteq [n]\\1\le |s|\le k}} \theta_s\, Z_s\Bigr)\,
H^{\otimes n},
\label{eq:iqp_unitary}
\end{equation}
where $[n]:=\{1,2,\ldots,n\}$, and for any subset $s\subseteq[n]$ we define the
multi-qubit Pauli strings
\[
X_s := \prod_{j\in s} X_j,
\qquad
Z_s := \prod_{j\in s} Z_j .
\]
The $k$-local restriction is enforced by setting $\theta_s=0$ for all subsets
with $|s|>k$.
Since $[X_s,X_{s'}]=0$ for all $s,s'\subseteq[n]$, the family $\{X_s\}$ is
simultaneously diagonalizable in the $X$-eigenbasis, and $U_{\mathrm{IQP}}$ is
diagonal in that basis. The associated IQP sampling task is to prepare $|0^n\rangle$, apply
$U_{\mathrm{IQP}}$, and measure in the computational
basis. 
In the ideal noiseless model, approximating this distribution is widely believed to be classically intractable (under standard assumptions used in the classical hardness literature \cite{bremner_classical_2010, Bremner_2016}).

\subsection{Noise model with interleaved Pauli noise}

As in \cite{rajakumar2025polynomial}, we consider an IQP circuit in which, after each unit of circuit depth, each qubit undergoes a single-qubit Pauli channel with probabilities $(p_I, p_X, p_Y, p_Z)$. A key scalar parameter controlling the transition is defined as 
\begin{equation}
    p \;:=\; p_Z \;+\; \min(p_X,\,p_Y).
\end{equation}
Intuitively, $p$ captures the part of the noise that most directly ``classicalizes” qubits relative to the $X$-diagonal structure and drives the breakup of long-range quantum correlations. 

\begin{figure*}[t]
    \centering
    \begin{subfigure}[b]{0.48\textwidth}
        \centering
        \includegraphics[width=\textwidth]{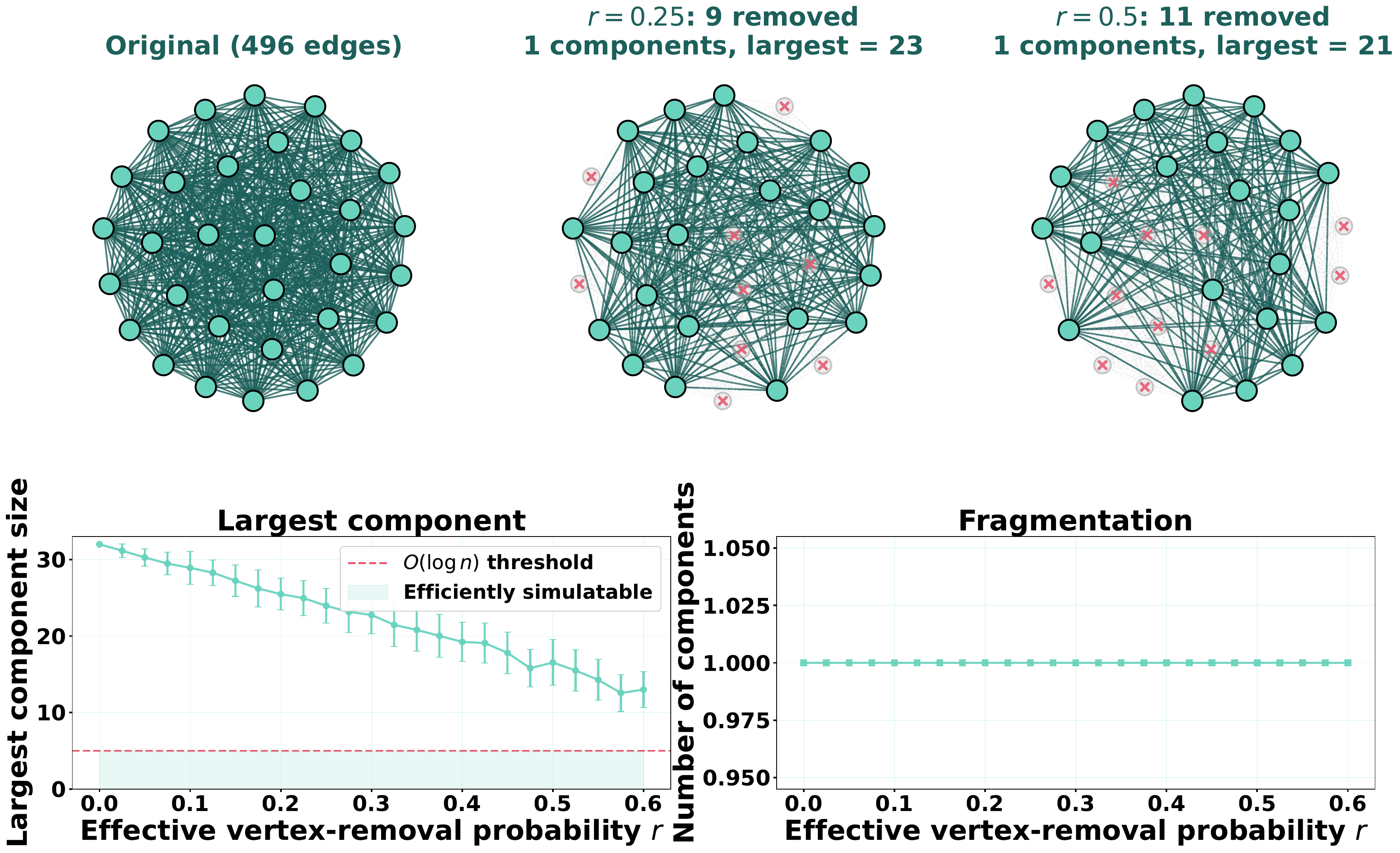}
        \caption{Dense ($n{=}32$, all $\binom{32}{2}$ interactions)}
    \end{subfigure}\hfill
    \begin{subfigure}[b]{0.48\textwidth}
        \centering
        \includegraphics[width=\textwidth]{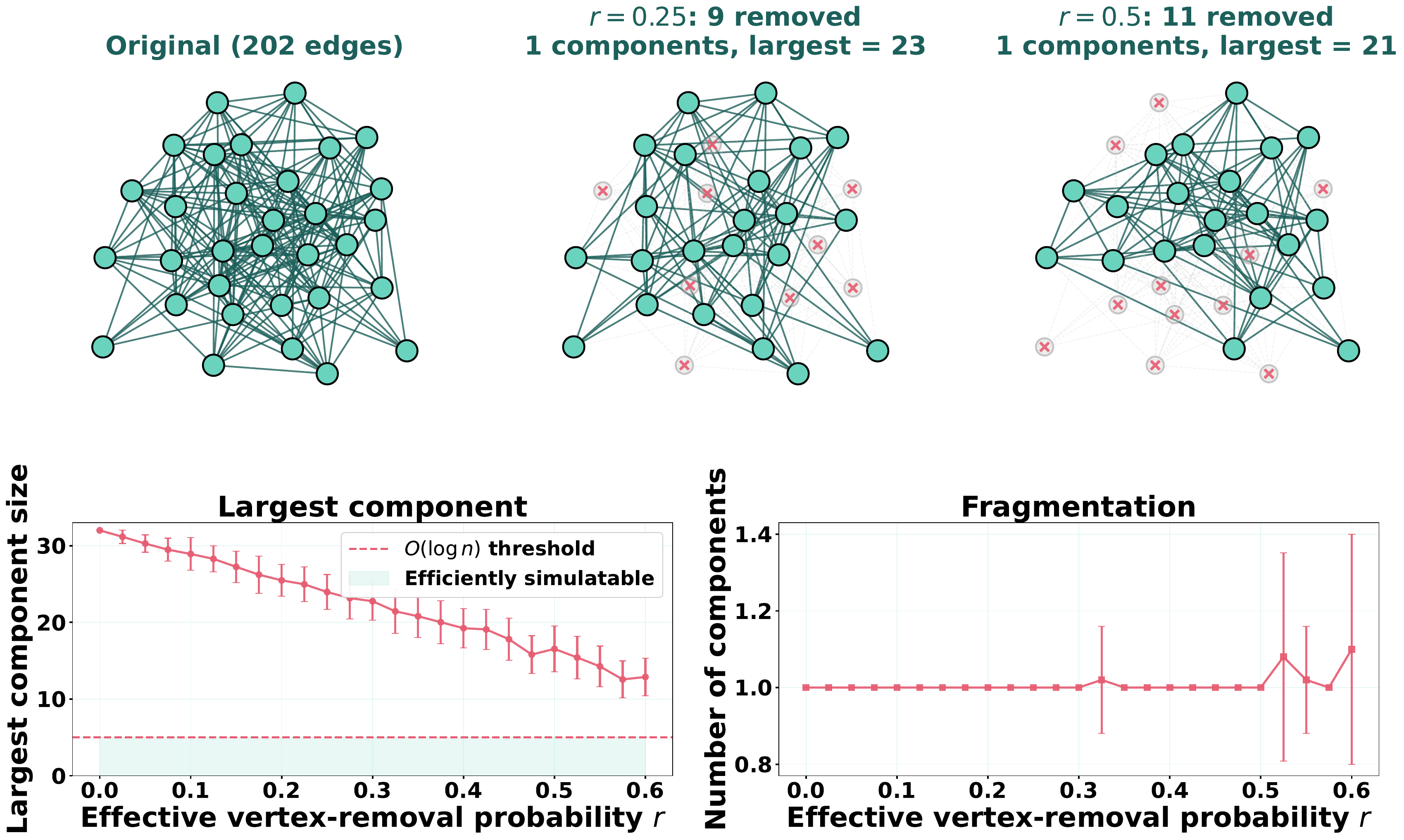}
        \caption{Sparse ($n{=}32$, density~$0.4$)}
    \end{subfigure}

    \vspace{0.3em}

    \begin{subfigure}[b]{0.48\textwidth}
        \centering
        \includegraphics[width=\textwidth]{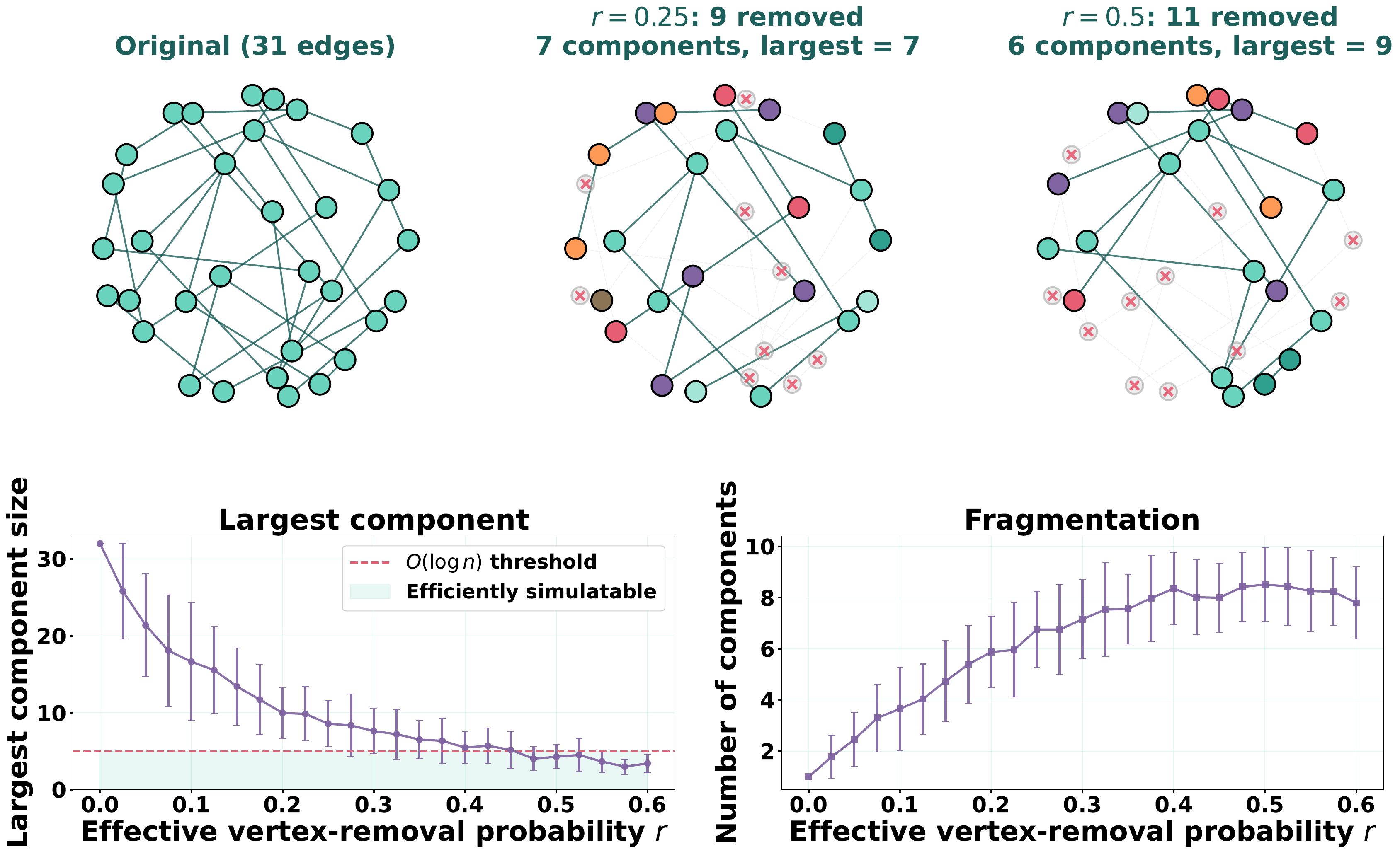}
        \caption{Local ($n{=}32$, nearest-neighbour)}
    \end{subfigure}\hfill
    \begin{subfigure}[b]{0.48\textwidth}
        \centering
        \includegraphics[width=\textwidth]{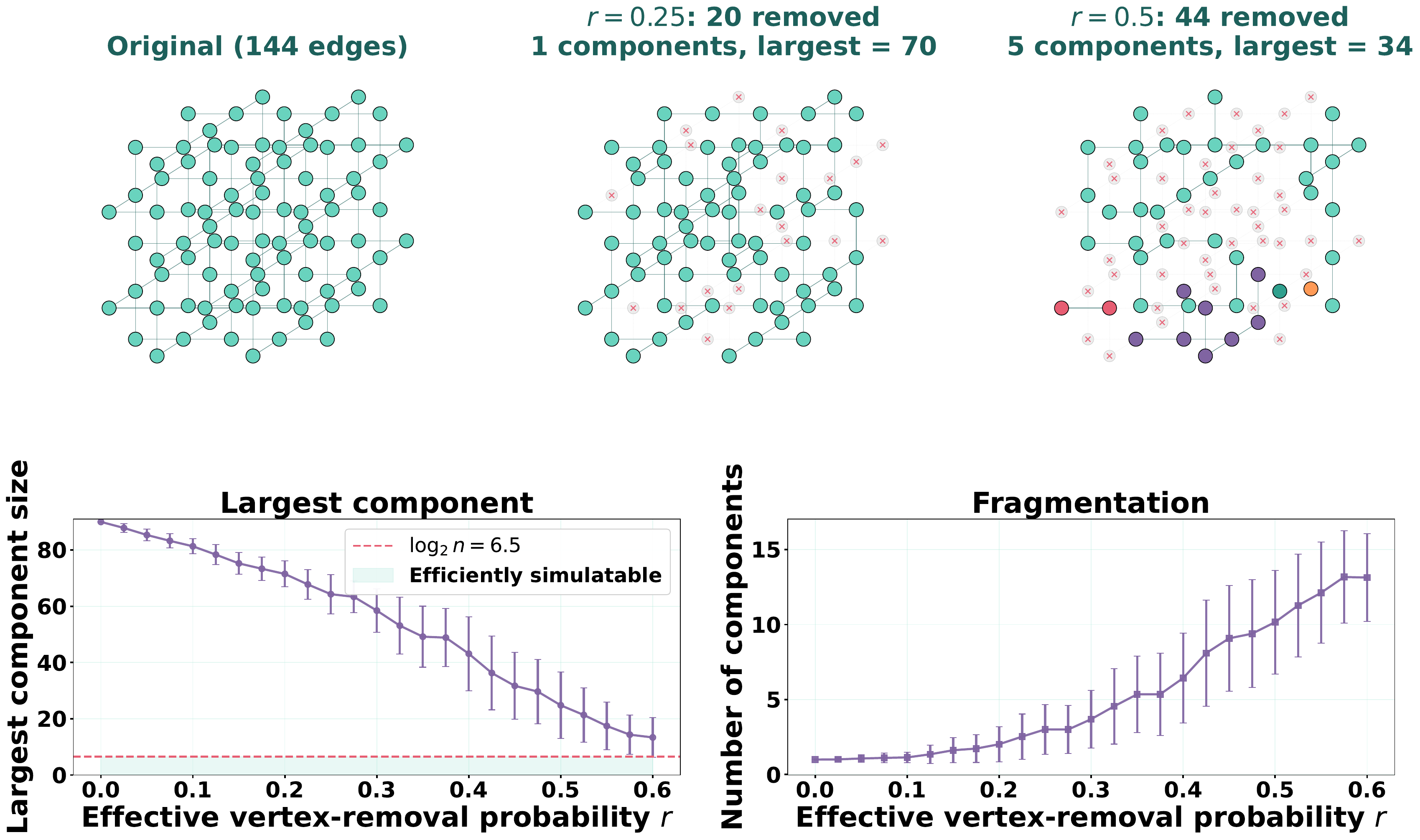}
        \caption{RHG lattice ($n{=}90$, 144 interactions)}
    \end{subfigure}

    \caption{Percolation-based fragmentation for four $k{=}2$ IQP interaction
        patterns. Panels (a)--(c) use $n{=}32$, while panel (d) uses an $n{=}90$
        RHG lattice. Each panel shows representative realizations and component
        statistics over 50--100 trials. The dashed line marks the $O(\log n)$
        component-size scale associated with efficient classical simulation.}
    \label{fig:percolation}
\end{figure*}

The proof strategy can be understood through an interaction graph whose vertices are qubits and whose edges represent entangling interactions induced by the $k$-local terms. Under the interleaved noise process, qubits are effectively disentangled with some probability, deleting incident edges and fragmenting the graph into small connected components. Fig.~\ref{fig:percolation} illustrates this process using the removal probability $r$, with $n=32$ for the dense, sparse, and local patterns and $n=90$ for the RHG lattice. Once all connected components are of size $O(\log n)$, each can be simulated independently and the overall noisy IQP distribution becomes efficiently samplable classically~\cite{markov2008simulating, de2025universal}.


\subsection{Depth criterion for the simulatable regime}
\label{sec:depth_cri}
Above a certain depth threshold, the fragmentation mechanism guarantees (with high probability) that the relevant interaction components remain small, enabling efficient simulation. For IQP circuits built from $k$-local diagonal gates under the Pauli noise model above, the sufficient condition for efficient classical simulation can be expressed (suppressing constants and low-order terms) as a depth bound of the form \cite{rajakumar2025polynomial}
\begin{equation}
\label{eq:critical_depth}
d_\ast(p,k)
=
O\!\left(
\frac{\log(k/p)}{p}
\right).
\end{equation}
Here $p$ is defined in Eq.~(2), and $d_\ast$ follows the notation of
Ref.~\cite{rajakumar2025polynomial}. If $d>d_\ast(p,k)$, there exists a
randomized classical algorithm that samples from the noisy circuit in
polynomial time. Higher noise lowers the critical-depth scale, while the
locality $k$ enters through a logarithmic factor. Conversely,
$d<d_\ast(p,k)$ does not establish classical hardness; it only means that
this sufficient simulatability condition does not apply.

The theorem is stated for an abstract noisy IQP circuit, but it already suggests the architectural lever we exploit later: any compilation step that increases depth and/or increases per-layer noise pushes the implementation toward simulatability. In limited-connectivity architectures, routing (SWAP networks), additional two-qubit gates, and idle-time exposure can all increase $(p, d)$. Fully connected architectures reduce routing overhead, potentially keeping the same abstract IQP family in a regime where the noisy circuit remains outside the known polynomial-time simulatability conditions.

\section{Methods}
\label{sec:methods}
\subsection{Connectivity-induced simulatability shift}
\label{sec:ciss}
We formalize how hardware connectivity changes the effective position of an IQP implementation relative to the known noisy-IQP simulatability boundary by explicitly separating an abstract circuit family from its compiled physical realizations.

Throughout this section, $D$ denotes the scheduled depth of a compiled circuit, measured in circuit-depth consistent with the noise insertion points of the noisy-IQP model. We use $p$ for the corresponding abstract effective noise strength and $H$ for a given hardware graph. These quantities are later instantiated as $(D_H, p_{\mathrm{eff}}(H))$ for compiled implementations on hardware graph $H$.
In our framework, compilation moves an instance in the $(D,p)$ plane relative to the critical-depth boundary $D_\ast(p,k)$ established in noisy-IQP analyses of \cite{rajakumar2025polynomial}. Noise is treated as effectively homogeneous and summarized through the pair $(D_H, p_{\mathrm{eff}}(H))$, allowing the implementation to be positioned relative to the known depth-noise phase boundary.

Let $C$ denote an $n$-qubit $k$-local IQP circuit instance specified by a set of commuting Pauli strings $\{X_s: 1\le |s|\le k\}$ with angles $\{\theta_s\}$ (cf. Eq.~\eqref{eq:iqp_unitary}). The circuit induces an interaction hypergraph $G_C$ over $[n]$, where each $k'$-body term $(k'\le k)$ corresponds to a hyperedge on the participating qubits.
We define $D_{\mathrm{FC}}$ as the scheduled depth required to implement $C$
assuming a fully connected hardware graph, i.e., any pair of qubits can be acted on
directly without routing (SWAP insertion). Importantly, this assumption concerns
only qubit connectivity: each $k$-local term in Eq.~\eqref{eq:iqp_unitary} is
still synthesized using the same fixed native gate set (single- and two-qubit
gates) as in the hardware-constrained setting.

Let $H$ denote a hardware connectivity graph on the same $n$ qubits (or on $N\ge n$ physical qubits with an injection map, which we suppress for clarity). In particular, we use $H_{\mathrm{FC}}$ for a fully connected graph, $H_{\mathrm{Dev}}$ for a device-specific hardware graph, and $H_{\mathrm{2D}}$ for a 2D nearest-neighbor grid.

Given a hardware graph $H$, compilation produces a physical circuit
$\widetilde{C}_H := \mathrm{Compile}(C;H)$,
which implements the same ideal unitary (up to permissible rewrites) using the native gate set and routing operations (e.g., SWAP networks) required by $H$. We denote by $D_H$ the resulting physical scheduled depth of $\widetilde{C}_H$ in units of circuit depth (or, more generally, in units consistent with the noise insertion points used in Sec.~\ref{sec:ptsiqp}).
The connectivity-induced depth overhead is then
\begin{equation}
\label{eq:delta_depth}
    \Delta D(H) \;:=\; D_H - D_{\mathrm{FC}} \;\ge\; 0.
\end{equation}
By construction, we expect
\begin{equation}
\Delta D(H_{\mathrm{FC}})\;\le\;\Delta D(H_{\mathrm{Dev}}),\ \Delta D(H_{\mathrm{2D}}),
\end{equation}
since full connectivity minimizes (and often eliminates) routing overhead for nonlocal interactions.

The noisy-IQP result reviewed in Eq.~\ref{eq:critical_depth} provides a noise-dependent critical depth scale $D_\ast(p,k)$ such that circuits exceeding this scale fall into an efficiently classically simulatable regime (up to the theorem’s approximation guarantees). We therefore interpret $\Delta D(H)$ as shifting the implemented circuit along the ``depth axis'' toward simulatability.

To make this quantitative in the global phase-diagram picture, we define a simulatability margin under hardware $H$ as
\begin{equation}
\label{eq:sim_margin}
m(H) \;:=\; D_\ast\!\bigl(p_{\mathrm{eff}}(H),k\bigr) \;-\; D_H,
\end{equation}
where $p_{\mathrm{eff}}(H)$ is an effective per-circuit-depth noise parameter for the compiled circuit $\widetilde{C}_H$. For $D_\ast\!\bigl(p_{\mathrm{eff}}(H),k\bigr)$ see Eq.~\eqref{eq:critical_depth}. Positive margin $m(H)>0$ indicates the implementation remains on the ``non-certified'' side of the known simulatability condition, whereas negative margin $m(H)<0$ indicates the implementation is predicted to lie in the simulatable regime.

Finally, we define the connectivity-induced simulatability shift relative to the ideal fully connected reference schedule as
\begin{equation}
\label{eq:transition}
S(H) \;:=\; \bigl(D_{\mathrm{FC}},p_{\mathrm{ideal}}\bigr)\ \mapsto\ \bigl(D_H, p_{\mathrm{eff}}(H)\bigr),    
\end{equation}
where $p_{\mathrm{ideal}}$ denotes the idealized reference noise level,
taken as zero in the noise-free baseline. Thus, $S(H)$ is the architecture-dependent map from an abstract instance to its effective operating point in the $(p,D)$ phase diagram. In our setting, the primary effect is an upward shift in depth by $\Delta D(H)$ (and typically a rightward shift in effective noise as well), with $H_{\mathrm{FC}}$ producing the smallest shift among the considered architectures.
We next quantify $D_H$ analytically for a representative sparse 2D grid geometry, which will later be used to evaluate the global margin $m(H)$ and the implied noise-budget ratios.

\subsection{Theoretical Estimation of \texorpdfstring{$D_H$}{DH} on a 2D Grid}
\label{sec:theory_DH_2D}

We next estimate the compiled depth $D_H$ for an $n$-qubit IQP instance mapped to a two-dimensional nearest-neighbor grid graph $H_{\mathrm{2D}}$, such as an $\sqrt{n}\times\sqrt{n}$ lattice. The objective is not to predict compiler-optimal constants, but rather to quantify how limited connectivity produces an upward shift in compiled depth and, consequently, reduces the simulatability margin defined by the critical-depth criterion in Eq.~\eqref{eq:critical_depth}.

\paragraph{Setup}
We assume a fixed native gate set with arbitrary one-qubit gates and a native
two-qubit entangling gate between adjacent vertices of $H_{\mathrm{2D}}$. Let $D_{\mathrm{FC}}$ be the scheduled compiled depth of an ideal IQP circuit $C$ under
full connectivity (no routing SWAPs), as defined in Sec.~\ref{sec:ciss}.
Compiling $C$ to the 2D grid produces a physical circuit
$\widetilde{C}_{\mathrm{2D}}$ with depth $D_{\mathrm{2D}}:=D_{H_{\mathrm{2D}}}$
and overhead $\Delta D_{\mathrm{2D}}:=D_{\mathrm{2D}}-D_{\mathrm{FC}}$.

We write $\mathrm{dist}_{\mathrm{2D}}(u,v)$ for the graph distance (Manhattan
distance) between two qubits $u,v$ on the grid.

\subsubsection*{(1) Random-pair interaction demand: distance-based lower bound}

A convenient abstraction is to view each two-qubit entangling operation in the
chosen 1q/2q decomposition of $C$ as a demand to apply an entangling gate on a
pair $(u,v)$. If $(u,v)$ is not an edge of $H_{\mathrm{2D}}$, routing is required
to bring the two qubits adjacent.

A simple distance-based lower bound follows from the fact that
each SWAP can reduce the distance between a target pair by at most $1$:
implementing a two-qubit gate on $(u,v)$ requires at least
$\mathrm{dist}_{\mathrm{2D}}(u,v)-1$ adjacency-creating moves.\footnote{Depending
on the routing model, the constant prefactor can vary (e.g., whether both qubits
move), but the distance scaling applies within this explicit state-routing model.} Summing over all
two-qubit entangling operations yields a coarse lower bound on routing cost and
hence on depth:
\begin{equation}
D_{\mathrm{2D}}
\;\gtrsim\;
D_{\mathrm{FC}}
\;+\;
\frac{1}{\Lambda}\sum_{(u,v)\in \mathcal{E}}
\bigl(\mathrm{dist}_{\mathrm{2D}}(u,v)-1\bigr),
\label{eq:DH_2D_dist_lower}
\end{equation}
where $\mathcal{E}$ is the multiset of two-qubit interaction pairs induced by
the chosen decomposition of $C$, and $\Lambda$ is the maximum number of
distance-reducing SWAP ``moves'' that can be executed in parallel per depth
(on a grid, $\Lambda=\Theta(n)$ in the best case).\footnote{Eq.~\eqref{eq:DH_2D_dist_lower}
is intentionally stated as an order-wise bound: it formalizes that large typical
distances force nontrivial depth overhead even under aggressive parallel
routing.}

For interaction demands that are effectively \emph{random} over pairs, which is
a reasonable proxy when the IQP instance induces near-complete mixing over
qubit pairs after decomposition, the typical grid distance satisfies
$\mathbb{E}[\mathrm{dist}_{\mathrm{2D}}(u,v)]=\Theta(\sqrt{n})$ for uniformly
random $u,v$ \cite{cowtan2019qubit}. Consequently, the average per-gate routing requirement
scales as $\Theta(\sqrt{n})$, implying a multiplicative depth overhead that can
grow with $\sqrt{n}$ if many long-range pairs must be realized.

\subsubsection*{(2) Exploiting $k$-local structure: support diameter lower bound}

We now leverage the fact that IQP circuits are $k$-local in the sense of
Eq.~\eqref{eq:iqp_unitary}, each generator term $\exp(i\theta_s X_s)$ involves
only $|s|\le k$ qubits. Under a standard 1q/2q synthesis, each $k'$-body term
($k'\le k$) can be implemented using $O(k')$ two-qubit entangling gates, e.g.,
by computing parities onto one qubit in $s$ (CNOT tree), applying a one-qubit
phase rotation, and uncomputing. Thus, for constant $k$, each term expands to
$O(k)$ two-qubit gates, and the routing overhead is governed by the geometry of
the participating qubits on the grid.

Let $\mathrm{diam}_{\mathrm{2D}}(s):=\max_{u,v\in s}\mathrm{dist}_{\mathrm{2D}}(u,v)$
denote the diameter of the subset $s$ under the grid metric. Any 2D
implementation of a multi-qubit interaction supported on $s$ must communicate
information across this diameter, which induces a depth cost at least linear in
$\mathrm{diam}_{\mathrm{2D}}(s)$. In particular, there exists a constant $c>0$
such that the two-qubit depth required to realize a single $k'$-local term
obeys
\begin{equation}
D_{\mathrm{2D}}(s)
\;\ge\;
c\,\mathrm{diam}_{\mathrm{2D}}(s),
\label{eq:term_depth_lower}
\end{equation}
up to additive $O(1)$ terms that depend only on the synthesis approach.
Therefore, for an IQP instance with term set $\mathcal{S}$ and a partition $\{\mathcal{S}_t\}_{t=1}^{D_{\mathrm{FC}}}$ under complete
connectivity, a conservative lower bound is
\begin{equation}
D_{\mathrm{2D}}
\;\gtrsim\;
\sum_{t=1}^{D_{\mathrm{FC}}}
\max_{s\in \mathcal{S}_t}\mathrm{diam}_{\mathrm{2D}}(s).
\label{eq:DH_2D_klocal_lower}
\end{equation}
For random $k$-tuples $s$ placed on an $\sqrt{n}\times\sqrt{n}$ grid,
$\mathbb{E}[\mathrm{diam}_{\mathrm{2D}}(s)]=\Theta(\sqrt{n})$ even for constant
$k\ge 2$, so Eq.~\eqref{eq:DH_2D_klocal_lower} again predicts a nontrivial upward
depth shift unless placement restricts the supports $s$ to be geometrically
local.

\paragraph{Constructive upper bound (order-wise)}
Conversely, standard 2D routing results imply that arbitrary qubit permutations
can be implemented in $O(\sqrt{n})$ SWAP depth on a 2D grid. As a consequence, an
arbitrary complete-graph two-qubit layer (a matching) can be realized with an
$O(\sqrt{n})$ depth overhead in the worst case by (i) routing qubits to make the
desired pairs adjacent, (ii) applying all entangling gates in parallel, and
(iii) optionally routing back. Therefore, for circuits whose ideal schedule
consists of $D_{\mathrm{FC}}$ two-qubit layers, we obtain the order-wise
upper bound
\begin{equation}
D_{\mathrm{2D}}
\;\lesssim\;
O(\sqrt{n})\cdot D_{\mathrm{FC}},
\label{eq:DH_2D_upper}
\end{equation}
with constants depending on the routing scheme and whether restoration to the
original layout is required.
Equations~\eqref{eq:DH_2D_dist_lower}--\eqref{eq:DH_2D_upper}
describe conventional state routing, not universal bounds over equivalent
syntheses. Parity-flow and SWAP-less methods may achieve smaller depth
~\cite{connectivity-aware,swapless}.

\paragraph{Implication for the simulatability margin on a 2D grid}
Combining the depth estimate above with the effective-noise aggregation model in
Sec.~\ref{sec:ptsiqp} yields an estimated operating point
$(p_{\mathrm{eff}}(H_{\mathrm{2D}}),D_{\mathrm{2D}})$. We define the (bound-based)
simulatability margin on the 2D grid as
\begin{equation}
m_{\mathrm{2D}}
\;:=\;
D_\ast\!\bigl(p_{\mathrm{eff}}(H_{\mathrm{2D}}),k\bigr)
\;-\;
D_{\mathrm{2D}},
\label{eq:margin_2D}
\end{equation}
where $D_\ast(p,k)$ follows the critical-depth scaling in Eq.~\eqref{eq:critical_depth}.
Using the upper bound in Eq.~\eqref{eq:DH_2D_upper} gives a conservative
sufficient condition for entering the simulatable regime:
\begin{equation}
D_{\mathrm{2D}}
\;\gtrsim\;
D_\ast\!\bigl(p_{\mathrm{eff}}(H_{\mathrm{2D}}),k\bigr),
\label{eq:2D_sufficient_simulable}
\end{equation}
whereas the lower bounds in Eqs.~\eqref{eq:DH_2D_dist_lower}
and~\eqref{eq:DH_2D_klocal_lower} quantify when nonlocal interaction supports
necessarily induce a large upward shift in depth.

Because the simulatability theory defines $p$ for an abstract interleaved Pauli
channel, we do not interpret $p$ as a directly measured hardware gate error
rate. Instead, we use the phase boundary as a \emph{noise-budget constraint}. For depth $D$, define the largest noise strength below the
sufficient-simulation threshold as
\[
p_{\mathrm{req}}(D,k) := \sup\{p: D \le D_\ast(p,k)\}.
\]
For two architectures with depths $D_{\mathrm{FC}}$ and
$D_{\mathrm{S}}=D_{\mathrm{FC}}+\Delta D$ (sparse), monotonicity of $D_\ast$ in
$p$ implies $p_{\mathrm{req}}(D_{\mathrm{S}},k) \le p_{\mathrm{req}}(D_{\mathrm{FC}},k)$.
Moreover, using the scaling $D_\ast(p,k)=\Theta\!\bigl(p^{-1}\log(k/p)\bigr)$ (Eq.~\ref{eq:critical_depth}) yields the first-order ratio estimate
\begin{equation}
\frac{p_{\mathrm{req}}(D_{\mathrm{S}},k)}{p_{\mathrm{req}}(D_{\mathrm{FC}},k)}
\;=\;
O\!\left(\frac{D_{\mathrm{FC}}}{D_{\mathrm{S}}}\right)
\;=\;
O\!\left(\frac{D_{\mathrm{FC}}}{D_{\mathrm{FC}}+\Delta D}\right),
\label{eq:noise_budget_ratio}
\end{equation}
where the $O(\cdot)$ notation absorbs constant factors together with the slowly varying logarithmic correction arising from $\log(k/p)$. Because Eq.~\eqref{eq:noise_budget_ratio} depends on achieved depth, it also
applies to alternative routing and synthesis methods. Consequently, \textit{architectures with limited connectivity generally require a lower effective noise level to compensate for routing-induced depth inflation}, whereas fully connected architectures can tolerate higher effective noise, or larger instances, while preserving the same simulatability margin.
\section{Results}
\label{sec:results}
We now evaluate the connectivity-induced simulatability shift $S(H)$ empirically across seven experimental hardware models grounded in publicly available device topologies and error-rate proxies, validating the theoretical framework of Sec.~\ref{sec:methods} and quantifying the simulatability margin $m(H)$ for simulated noisy hardware devices. All experiments use $k{=}2$ (pairwise) interactions for the IQP, since this setting already includes nontrivial IQP families relevant to near-term quantum advantage and captures the interaction structures studied in this work.

\subsection{Experimental Evaluation of Simulatability Margins}
\label{sec:phase-diagrams}

\begin{figure}[t]
    \centering

    \begin{subfigure}[t]{0.44\columnwidth}
        \centering
        \includegraphics[width=\linewidth]{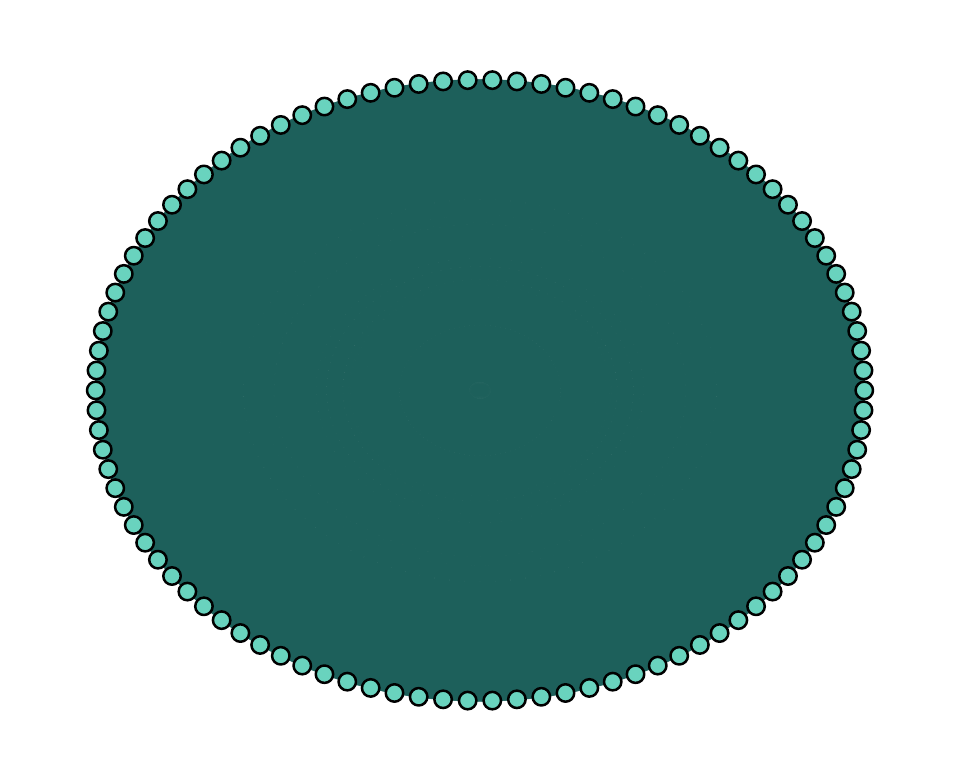}
        \caption{FC\_1 (98 qubits, all-to-all).}
    \end{subfigure}\hfill
    \begin{subfigure}[t]{0.44\columnwidth}
        \centering
        \includegraphics[width=\linewidth]{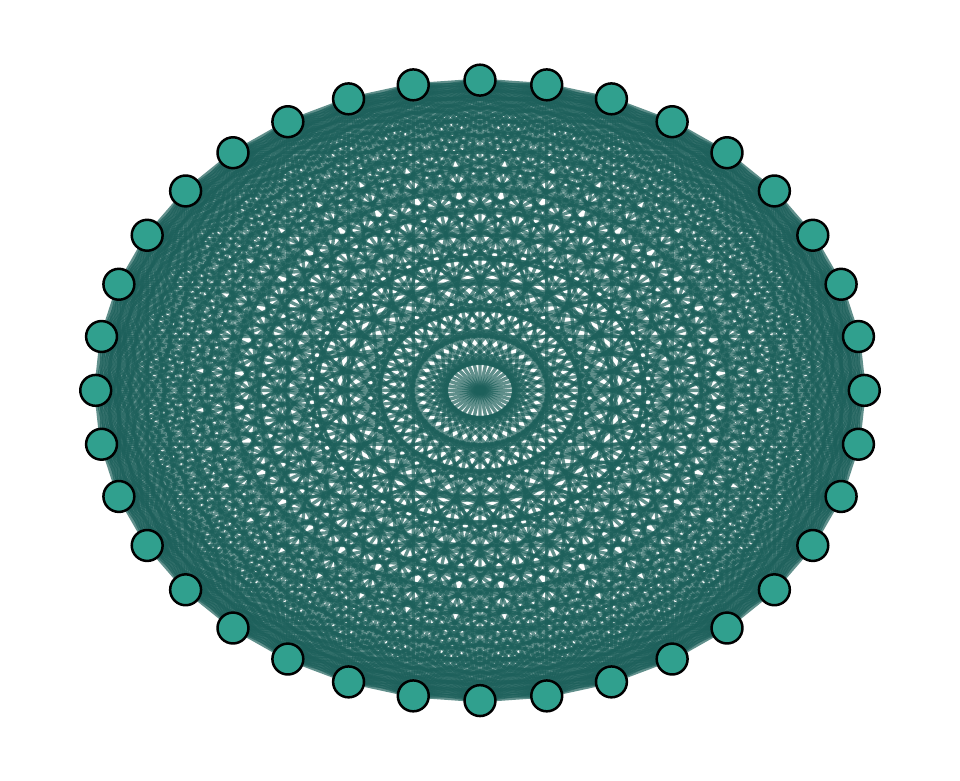}
        \caption{FC\_2 (36 qubits, all-to-all).}
    \end{subfigure}

    \vspace{1mm}

    \begin{subfigure}[t]{0.44\columnwidth}
        \centering
        \includegraphics[width=\linewidth]{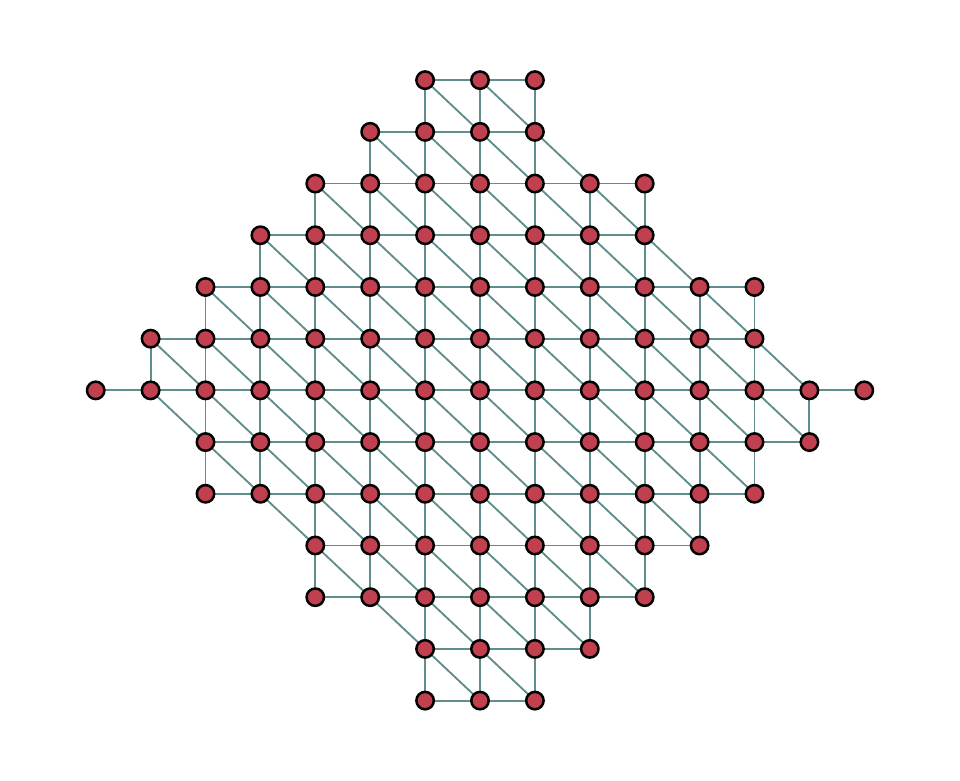}
        \caption{SC\_1 (105 qubits, tapered triangular lattice).}
    \end{subfigure}\hfill
    \begin{subfigure}[t]{0.44\columnwidth}
        \centering
        \includegraphics[width=\linewidth]{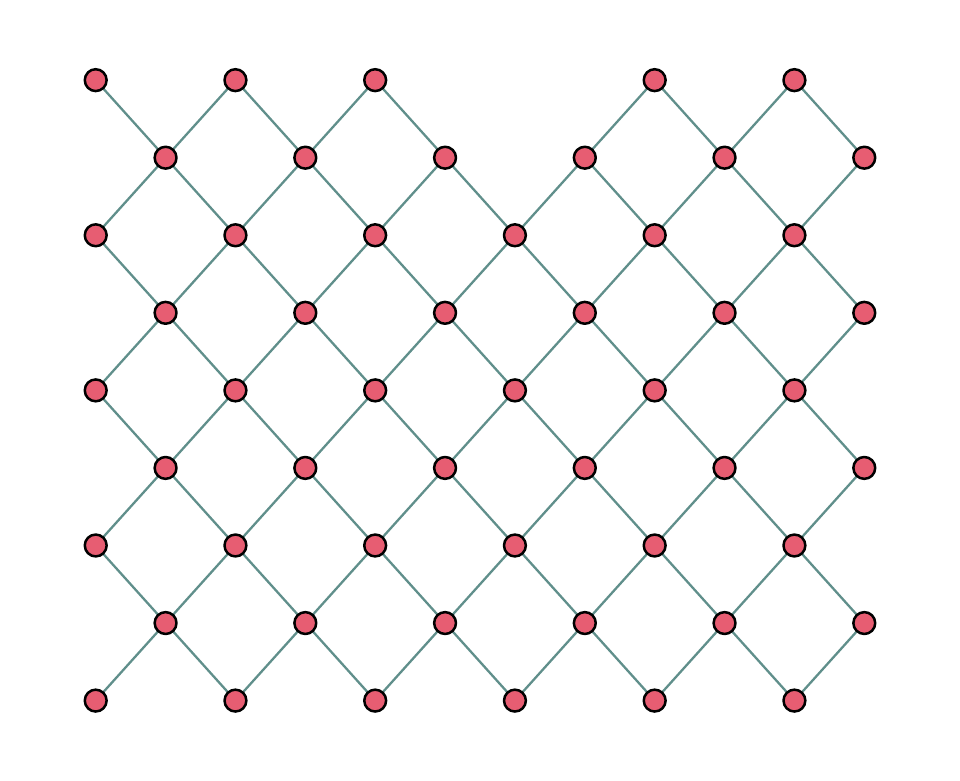}
        \caption{SC\_2 (53 qubits, staggered nearest-neighbor lattice).}
    \end{subfigure}

    \vspace{1mm}
    \begin{subfigure}[t]{0.44\columnwidth}
        \centering
        \includegraphics[width=\linewidth]{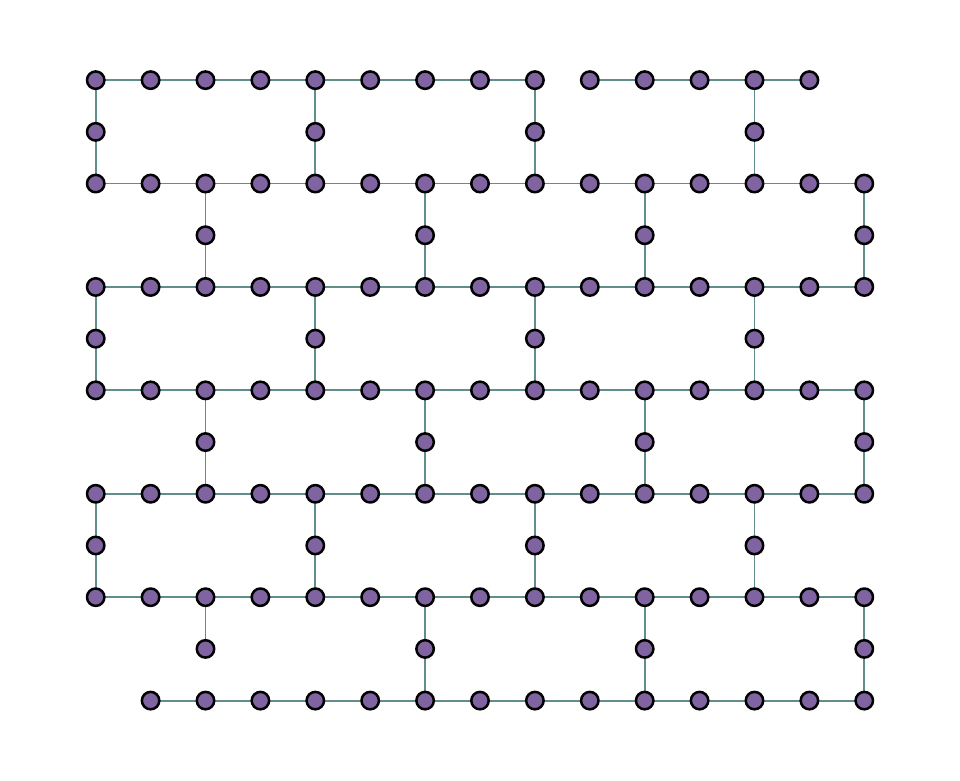}
        \caption{SC\_3 (127 qubits, heavy-hex connectivity).}
    \end{subfigure}\hfill
    \begin{subfigure}[t]{0.44\columnwidth}
        \centering
        \includegraphics[width=0.88\linewidth]{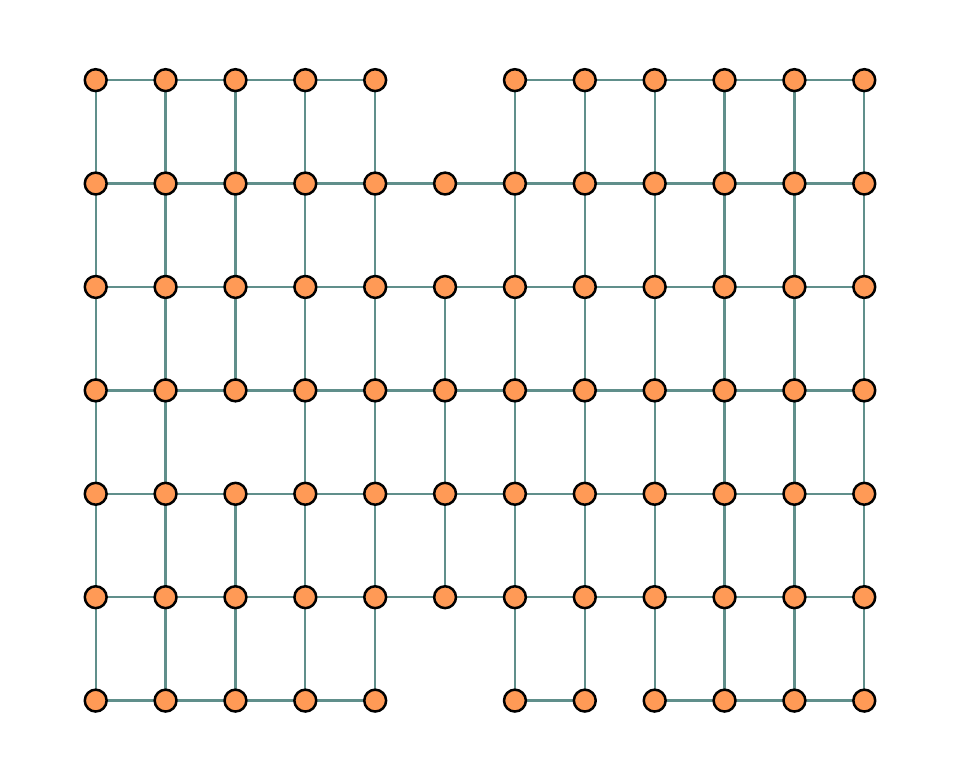}
        \caption{SC\_4 (82 qubits, defected rectangular grid).}
    \end{subfigure}

    \vspace{1mm}

    \begin{subfigure}[t]{0.44\columnwidth}
        \centering
        \includegraphics[width=\linewidth]{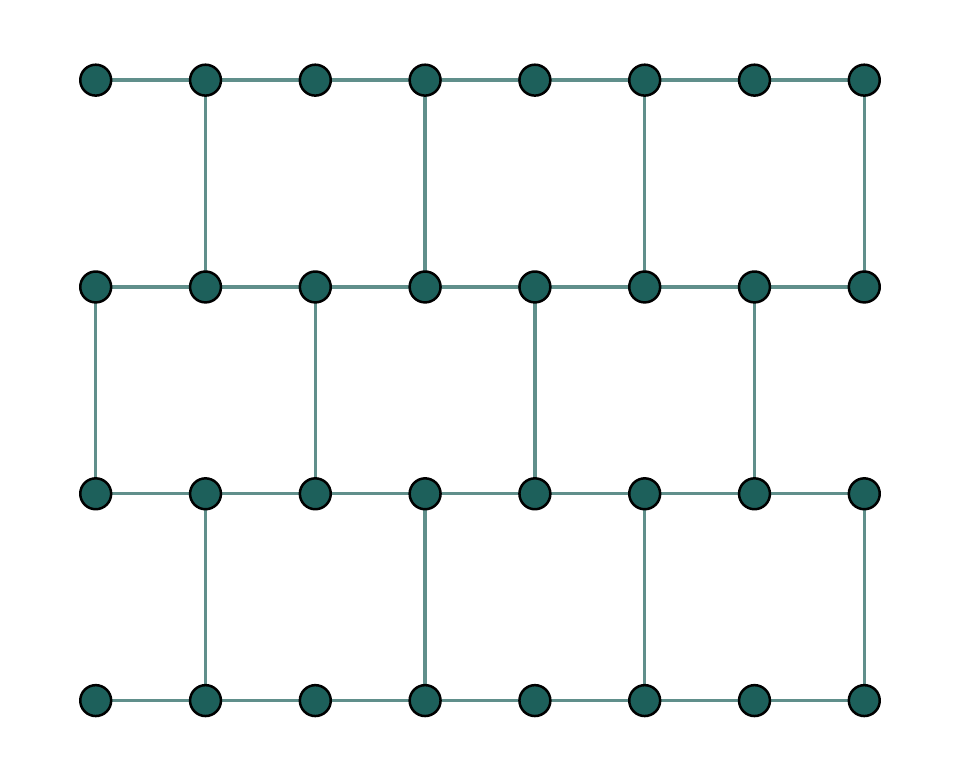}
        \caption{SC\_5 (32 qubits, approx.\ sparse near-planar lattice).}
    \end{subfigure}

    \caption{Native hardware connectivity graphs used in the routing benchmark. The models are ordered from fully connected trapped-ion-like architectures to more locality-constrained superconducting and planar-style layouts, illustrating the wide range of coupling structures that drive different routing overheads.}
    \label{fig:hardware_topologies}
\end{figure}

\subsubsection{Compilation Setup}
We compile $k{=}2$ IQP circuits across four interaction patterns:
\begin{itemize}
    \item \textbf{Dense}: all pairwise interactions on $n{=}32$ qubits,
    giving $\binom{32}{2}=496$ terms.
    \item \textbf{Sparse}: a random subset of the $n{=}32$ pairwise
    interactions at density~$0.4$.
    \item \textbf{Local}: nearest-neighbor interactions on an $n{=}32$
    qubit 1D chain, giving 31 terms.
    \item \textbf{RHG lattice}: interactions defined by a Raussendorf--Harrington--Goyal lattice~\cite{Raussendorf_2007}, the three-dimensional cluster-state topology used in topological measurement-based quantum computation.
\end{itemize}

For the dense, sparse, and local patterns, we use $n{=}32$ qubits in the phase-diagram study. For the RHG pattern, we additionally consider larger device-scale instances, since the associated commuting-circuit sampling task is believed to remain classically hard under standard complexity-theoretic assumptions while requiring only $O(n)$ two-qubit gates~\cite{Fujii2016NoisyCommuting}. Specifically, we compile RHG circuits at $n{=}90$ ($2{\times}2{\times}2$ cells, 144 edges) and $n{=}127$ ($2{\times}2{\times}3$ cells, 208 edges).

Raw ideal IQP circuits are generated in
\texttt{qiskit}~\cite{Qiskit} and compiled in
\texttt{pytket}~\cite{Sivarajah_2020}. For each hardware target $H$, we compute a hardware-constrained depth $D_H$ by routing on the coupling graph of $H$, rebasing to the target-family native gateset, and removing redundancies, together with a native fully connected baseline depth $D_{\mathrm{FC}}$ obtained by compiling the same ideal circuit to the same native gateset without connectivity constraints. Here, depth denotes the circuit depth of the final compiled native circuit after routing and rebasing. This isolates topology-induced routing overhead from native-gateset effects and yields $\Delta D(H)=D_H-D_{\mathrm{FC}}$ as in Eq.~\eqref{eq:delta_depth}.
The native gate families used in the current experiments are: FC\_1 $\{R_z,\mathrm{PhasedX},\mathrm{ZZPhase}\}$, FC\_2 $\{\mathrm{GPI},\mathrm{GPI2},\mathrm{ZZPhase}\}$, SC\_1/SC\_2 $\{R_z,\mathrm{PhasedX}\text{-type},\mathrm{CZ}\}$ as available in the \texttt{pytket} version used for these experiments, SC\_3 $\{R_z,\mathrm{SX},X,\mathrm{CX}\}$, SC\_4 $\{R_z,\mathrm{RX},\mathrm{ISWAP}\}$, and SC\_5 $\{R_z,\mathrm{SX},\mathrm{ECR}\}$.

The hardware connectivity graphs used in these simulations are shown in Fig.~\ref{fig:hardware_topologies}. Throughout, FC denotes fully connected hardware models and SC denotes sparse-connectivity hardware models. FC\_1, FC\_2, SC\_1, SC\_2, SC\_3, and SC\_4 are grounded on device layouts reconstructed from public vendor documentation, SDK coupling maps, or publicly released hardware diagrams. SC\_5 is the only exception: because we did not identify a public qubit-level coupling map for that processor, we model it with a sparse near-planar 32-qubit proxy architecture intended to capture the qualitative connectivity constraints of a generic superconducting device. All results in this section are obtained in classical simulation of these connectivity and native-gate models.

In this work, we approximate the effective noise parameter $p_{\mathrm{eff}}(H)$ with two-qubit gate error rates assigned from published or publicly available calibration proxies: FC\_1, $7.9 \times 10^{-4}$~\cite{ransford2025helios98qubittrappedionquantum}; FC\_2, $4.0 \times 10^{-3}$~\cite{ionqForteErrorRates}; SC\_1, $3.3 \times 10^{-3}$~\cite{googleWillowSpecSheet,google2024willow,google2025surfacecode}; SC\_2, $6.0 \times 10^{-3}$~\cite{arute2019quantum}; SC\_3, $1.146 \times 10^{-2}$~\cite{qiskit_fake_washington}; SC\_4, $1.94 \times 10^{-2}$~\cite{rigetti_ankaa3}; and SC\_5, $4.6 \times 10^{-2}$~\cite{ward2026echocrossresonancegate}. This proxy is used for comparative architecture-level analysis and should not be interpreted as a calibrated estimate of the abstract Pauli-noise parameter in the theorem.

\subsubsection{Phase Diagram Results}

\begin{figure*}[!t]
    \centering
    \begin{subfigure}[b]{0.42\textwidth}
        \centering
        \includegraphics[width=\textwidth]{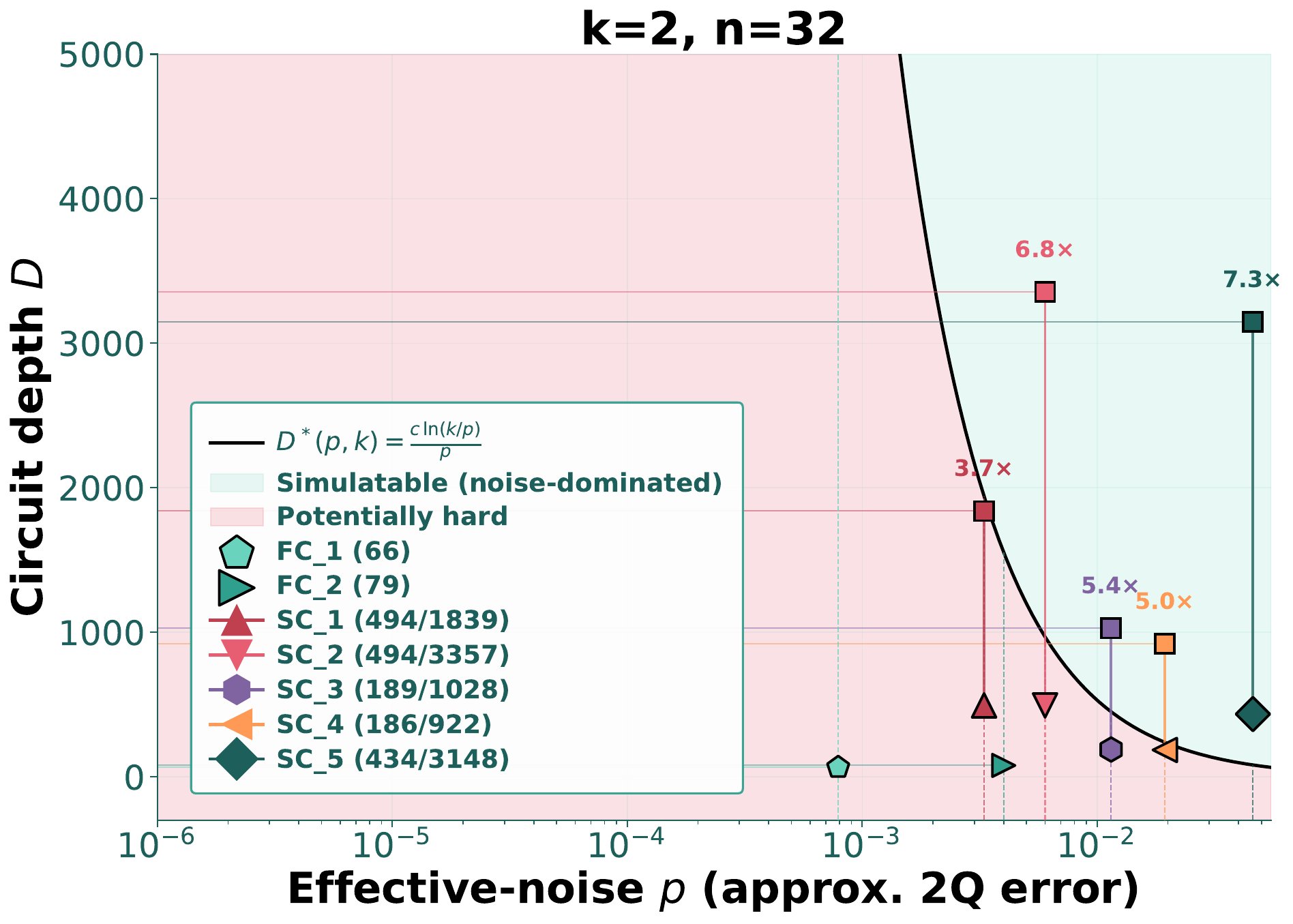}
        \caption{Dense (all $\binom{32}{2}=496$ interactions)}
        \label{fig:phase_dense}
    \end{subfigure}
    \begin{subfigure}[b]{0.42\textwidth}
        \centering
        \includegraphics[width=\textwidth]{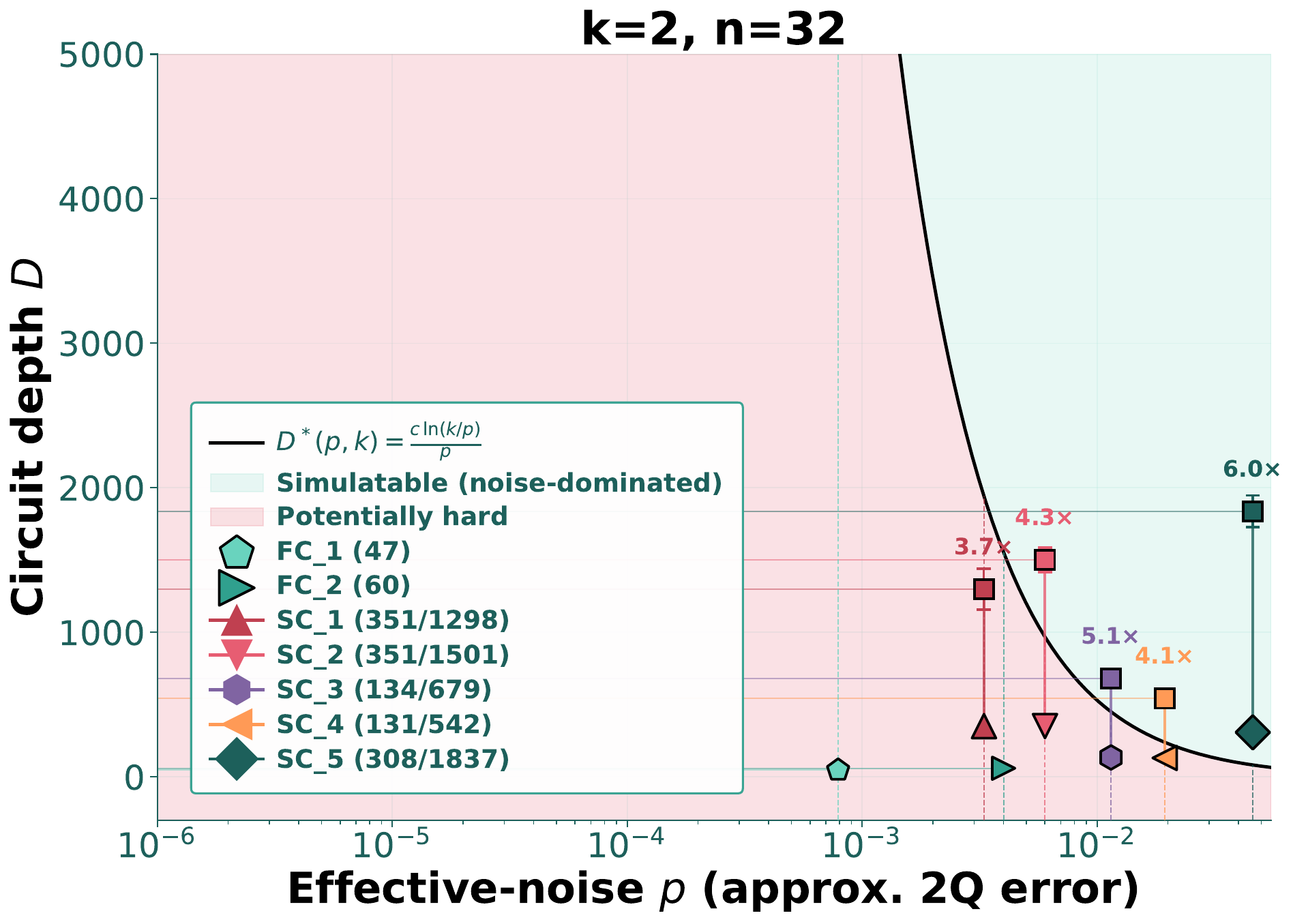}
        \caption{Sparse (random subset, density~$0.4$)}
        \label{fig:phase_sparse}
    \end{subfigure}
    
    \vspace{0.27cm}
    
    \begin{subfigure}[b]{0.42\textwidth}
        \centering
        \includegraphics[width=\textwidth]{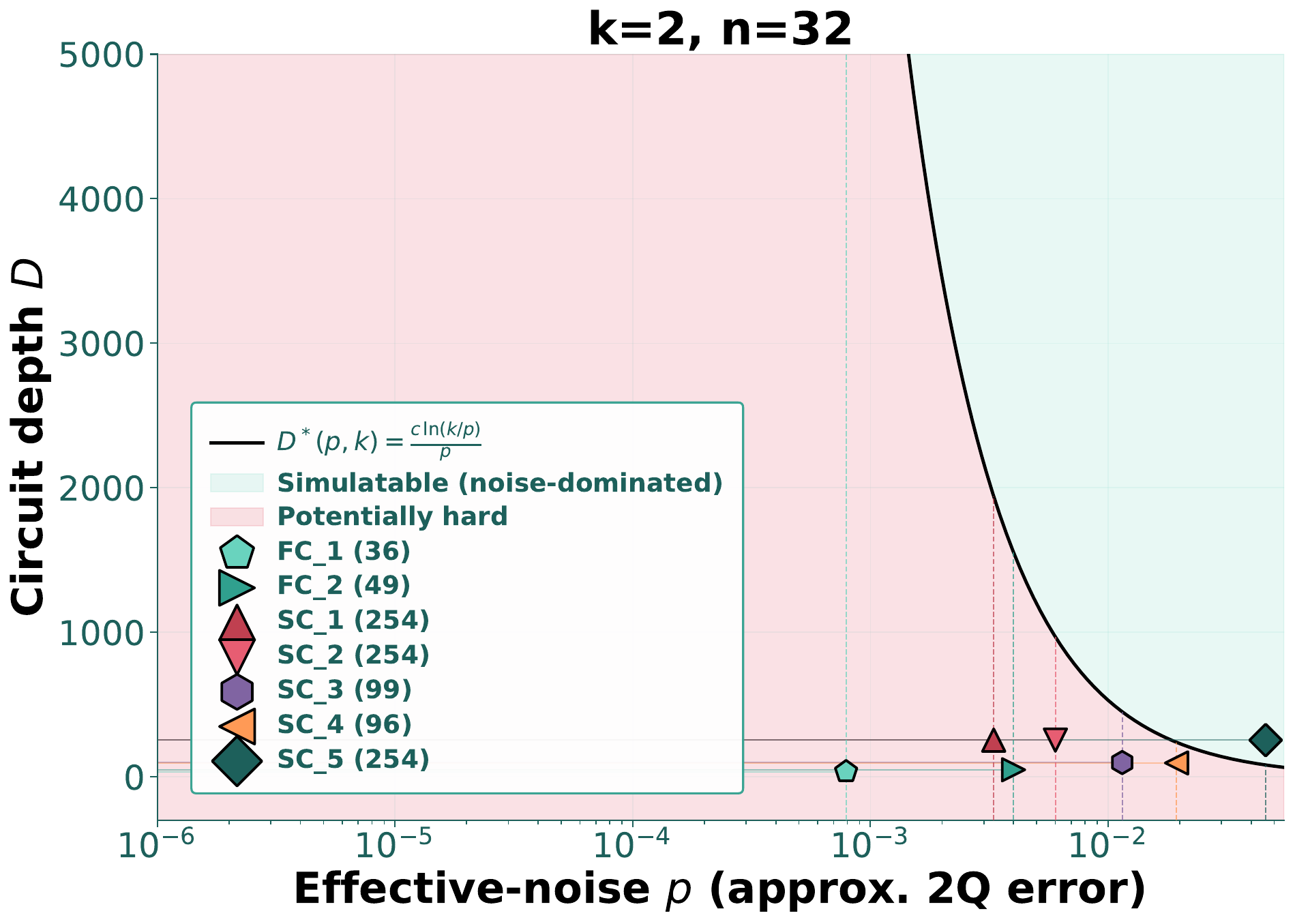}
        \caption{Local (nearest-neighbor, 31 interactions)}
        \label{fig:phase_local}
    \end{subfigure}
    \begin{subfigure}[b]{0.42\textwidth}
        \centering
        \includegraphics[width=\textwidth]{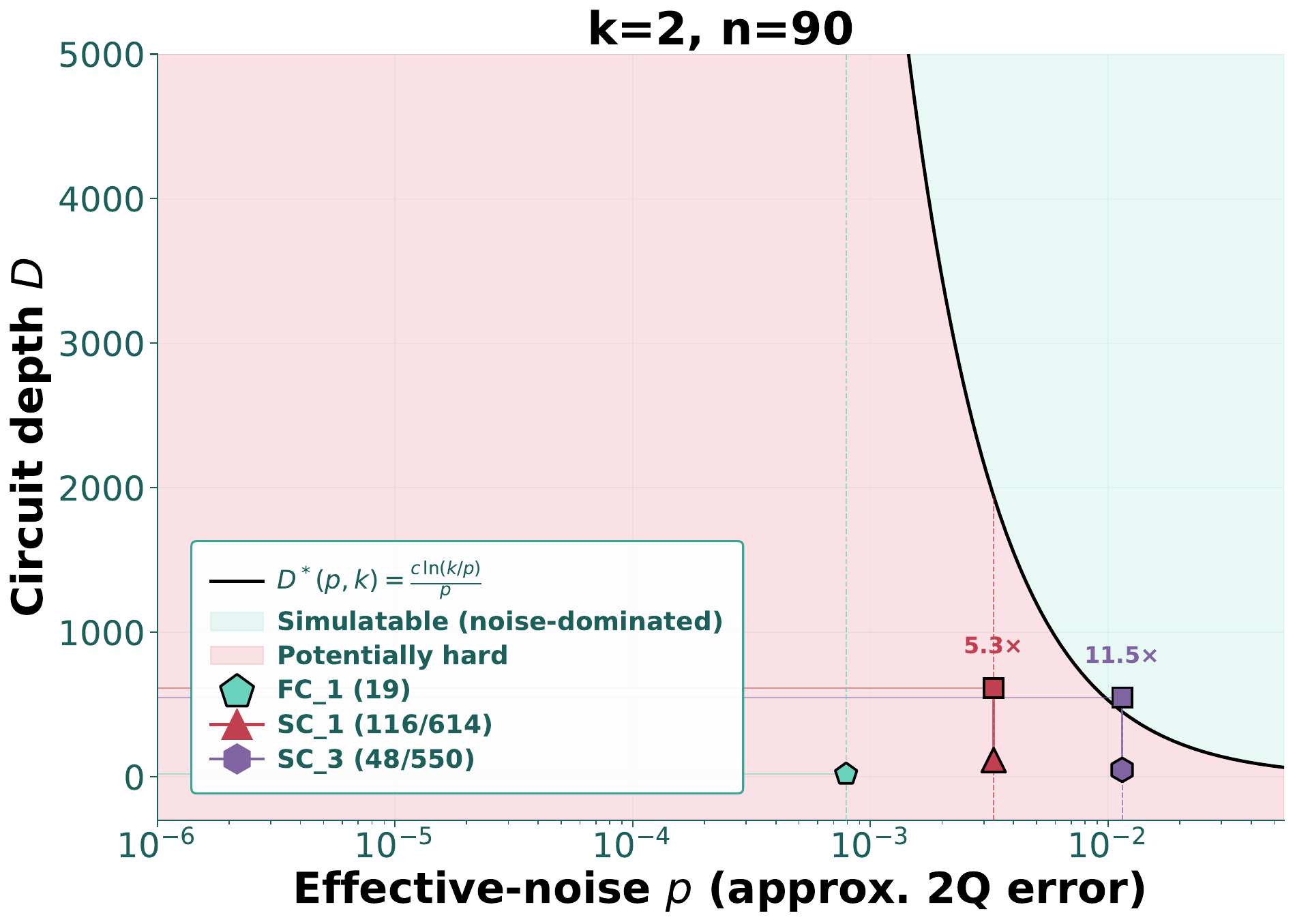}
        \caption{RHG lattice ($n{=}90$, 144 interactions)}
        \label{fig:phase_rhg}
    \end{subfigure}
    
    \caption{Phase diagrams for $k{=}2$ IQP circuits. The black curve shows
$D_\ast(p,k)=c\ln(k/p)/p$, with $c=1$. Points use
the assigned effective-noise proxy $p_{\mathrm{eff}}(H)$ and compiled depth
$D_H$. For sparse models, lower and upper markers denote $D_{\mathrm{FC}}$
and $D_H$, with vertical segments showing routing overhead. Under the proxy mapping and normalization, ``Simulatable'' satisfies the sufficient simulation
condition, whereas ``Potentially hard'' means only that this condition does
not apply. Values are means over twenty instances with
one-standard-deviation error bars. Panels (a)--(c) use $n{=}32$, and panel
(d) uses $n{=}90$.}
    \label{fig:phase_diagrams}
\end{figure*}

\begin{table}[h]
\centering
\footnotesize
\caption{Compiled depth $D_H$ for $k{=}2$ IQP circuits across four interaction patterns. Dense, Sparse, and Local use $n{=}32$; RHG uses $n{=}90$ (only hardware with $\ge 90$ qubits shown). Entries are means over twenty circuit instances per pattern (depths rounded to the nearest integer). Values in parentheses are the depth overhead $D_H/D_{\mathrm{FC}}$ computed from the means.}
\label{tab:phase_summary}
\setlength{\tabcolsep}{3pt}
\begin{tabular}{l@{\hskip 6pt}r@{\hskip 6pt}r@{\hskip 6pt}r@{\hskip 6pt}r}
\toprule
\textbf{Platform} & \textbf{Dense} & \textbf{Sparse} & \textbf{Local} & \textbf{RHG} \\
\midrule
FC\_1  & 66\,(1.0$\times$) & 47\,(1.0$\times$) & 36\,(1.0$\times$) & 19\,(1.0$\times$) \\
FC\_2  & 79\,(1.0$\times$) & 60\,(1.0$\times$) & 49\,(1.0$\times$) & --- \\
SC\_1  & 1839\,(3.7$\times$) & 1298\,(3.7$\times$) & 254\,(1.0$\times$) & 614\,(5.3$\times$) \\
SC\_2  & 3357\,(6.8$\times$) & 1501\,(4.3$\times$) & 254\,(1.0$\times$) & --- \\
SC\_3  & 1028\,(5.4$\times$) & 679\,(5.1$\times$) & 99\,(1.0$\times$) & 550\,(11.5$\times$) \\
SC\_4  & 922\,(5.0$\times$) & 542\,(4.1$\times$) & 96\,(1.0$\times$) & --- \\
SC\_5  & 3148\,(7.3$\times$) & 1837\,(6.0$\times$) & 254\,(1.0$\times$) & --- \\
\bottomrule
\end{tabular}
\end{table}

Fig.~\ref{fig:phase_diagrams} visualizes the operating points
$(p_{\mathrm{eff}}(H),D_H)$ relative to the instantiated boundary
$D_\ast(p,k)=c\ln(k/p)/p$ with $c=1$, and
Table~\ref{tab:phase_summary} summarizes the compiled depths.
Because Eq.~\eqref{eq:critical_depth} specifies only the asymptotic
scaling, the region labels and reported crossings are conditional on
this normalization and on the adopted effective-noise proxy.\footnote{The more
explicit thresholds derived in the appendix of Ref.~\cite{rajakumar2025polynomial} are lower
over the parameter range considered here, giving a less forgiving
classification for possible quantum advantage while leaving the measured
connectivity-induced depth shifts unchanged.}
For sparse-connectivity models, both the native fully connected baseline $D_{\mathrm{FC}}$ and the hardware-constrained depth $D_H$ are shown,
making the connectivity-induced vertical shift explicit. The reported values and variance profiles correspond to Qiskit logical-circuit construction followed by pytket native-gate compilation and the default pytket placement and routing strategy used in this study; because routing is heuristic, alternative mapping strategies may yield different variance profiles.

\paragraph{Dense interactions (worst case)}
The dense pattern includes all $\binom{32}{2}$ pairwise terms and produces the largest compilation cost. The two fully connected experimental models differ in absolute compiled depth: FC\_1 compiles to $D_H=66$ and FC\_2 to $D_H=79$, both with zero routing overhead and within the ``Potentially hard'' region. Among the locality-constrained models, SC\_4 is the shallowest at $D_H=922$ ($5.0\times$ overhead), while SC\_1, SC\_2, SC\_3, and SC\_5 reach $D_H=1839$, $3357$, $1028$, and $3148$, corresponding to $3.7\times$, $6.8\times$, $5.4\times$, and $7.3\times$ depth inflation over their native fully connected baselines.
Under the adopted effective-noise proxy, FC\_1, FC\_2, and SC\_1 remain in the ``Potentially hard'' region. The mean SC\_1 point lies close to but below the boundary, with $D_H=1839$ compared with $D_\ast\approx1942$ at its assigned error rate. For SC\_2, SC\_3, and SC\_4, connectivity increases the depths from $D_{\mathrm{FC}}=494$, $189$, and $186$ to $D_H=3357$, $1028$, and $922$, respectively, moving all three mean compiled implementations across the boundary into the sufficient simulatable region. SC\_5 lies in the sufficient simulatable region already at its native fully connected baseline $D_{\mathrm{FC}}=434$, and connectivity further increases its depth to $D_H=3148$.

\paragraph{Sparse and local interactions}
The sparse pattern reduces compiled depths relative to the dense case across
all constrained architectures: SC\_1 decreases from $1839$ to $1298$, SC\_2
from $3357$ to $1501$, SC\_5 from $3148$ to $1837$, SC\_4 from $922$ to $542$,
and SC\_3 from $1028$ to $679$. Under the adopted proxy mapping, FC\_1, FC\_2,
and SC\_1 remain in the ``Potentially hard'' region. Routing moves SC\_2,
SC\_3, and SC\_4 across the boundary into the sufficient simulatable region,
with mean depth inflation factors of $4.3\times$, $5.1\times$, and $4.1\times$,
respectively. SC\_5 lies in the sufficient simulatable region at both its
native fully connected baseline and its hardware-constrained depth, so no
additional connectivity-induced crossing occurs for the sparse pattern.

For the local pattern, all platforms satisfy $D_H=D_{\mathrm{FC}}$, so no
connectivity-induced routing overhead is incurred. SC\_5 is the only
local-pattern point in the sufficient simulatable region. Since routing
overhead vanishes in this case, its position is determined by its native
compilation depth and assigned noise proxy rather than by connectivity.

\paragraph{RHG lattice (device-scale)}
The RHG lattice provides the strongest theoretical setting: $k{=}2$ interactions on this 3D topology are a theoretically well-motivated candidate hard family~\cite{Fujii2016NoisyCommuting}.
At $n{=}90$ (Fig.~\ref{fig:phase_diagrams}d), FC\_1 achieves the shallowest depth, $D_H=19$, again with zero routing overhead. SC\_1 and SC\_3 compile to $D_H=614$ and $D_H=550$, corresponding to $5.3\times$ and $11.5\times$ overhead over their native fully connected baselines, respectively. Other approximated hardware models are excluded here because of the $n{=}90$ qubit size required. Under the adopted error proxy, FC\_1 and SC\_1 remain below the boundary; despite its larger absolute compiled depth, SC\_1 remains below its boundary value $D_\ast\approx1942$. In contrast, routing moves SC\_3 from $D_{\mathrm{FC}}=48$ to $D_H=550$, above its boundary value $D_\ast\approx450$ and into the sufficient simulatable region.
At $n{=}127$, only SC\_3 can accommodate the RHG instance among the hardware models considered, compiling to $D_H=1128$ with a $20.9\times$ depth overhead over its native fully connected baseline ($D_{\mathrm{FC}}=54$). The $n=127$ SC\_3 point also lies above the sufficient-simulation boundary. Thus, for the adopted proxy, the RHG interaction structure remains strongly routing-limited on SC\_3, while the $n=90$ SC\_1 implementation remains below the boundary despite a substantial connectivity-induced depth increase.

\subsection{Compilation Overhead Scaling}
\label{subsec:compilation_overhead}

\begin{figure}[htbp]
    \centering
    \begin{subfigure}[b]{0.44\textwidth}
        \includegraphics[width=\textwidth]{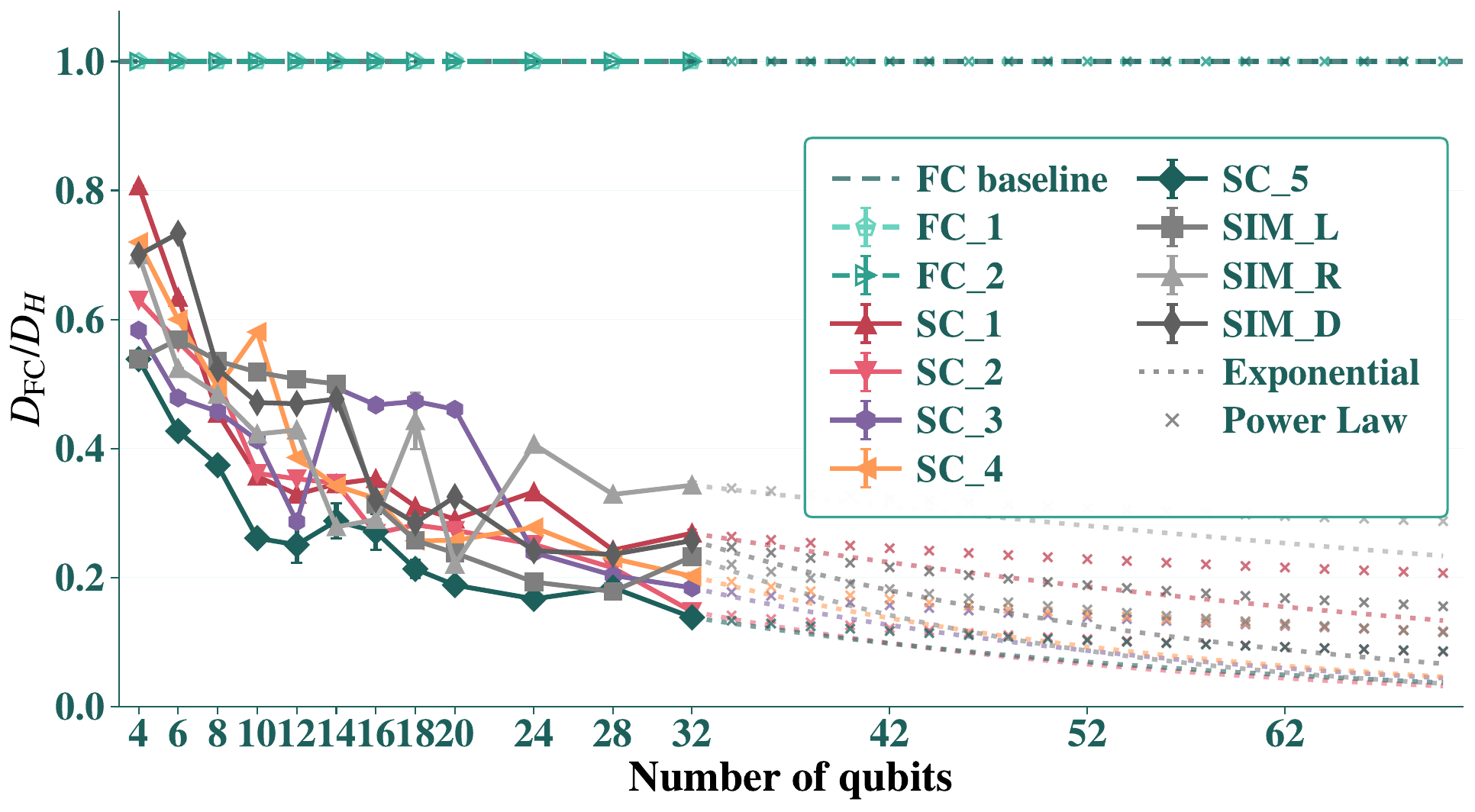}
        \caption{Dense: all $\binom{n}{2}$ interactions}
        \label{fig:ratio-dense}
    \end{subfigure}
    \hfill
    \begin{subfigure}[b]{0.44\textwidth}
        \includegraphics[width=\textwidth]{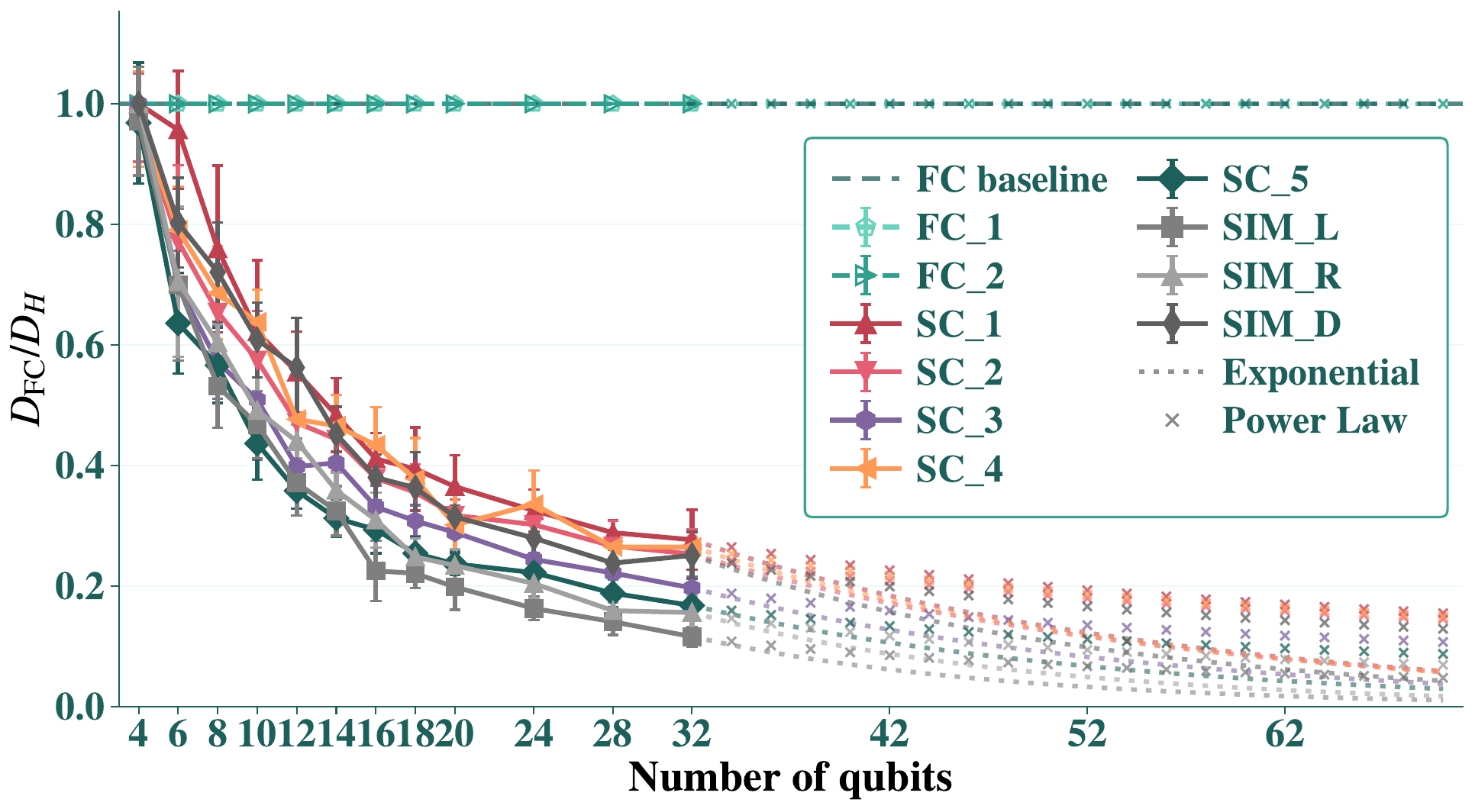}
        \caption{Sparse: random subset with density $0.4$}
        \label{fig:ratio-sparse}
    \end{subfigure}
    \hfill
    \begin{subfigure}[b]{0.44\textwidth}
        \includegraphics[width=\textwidth]{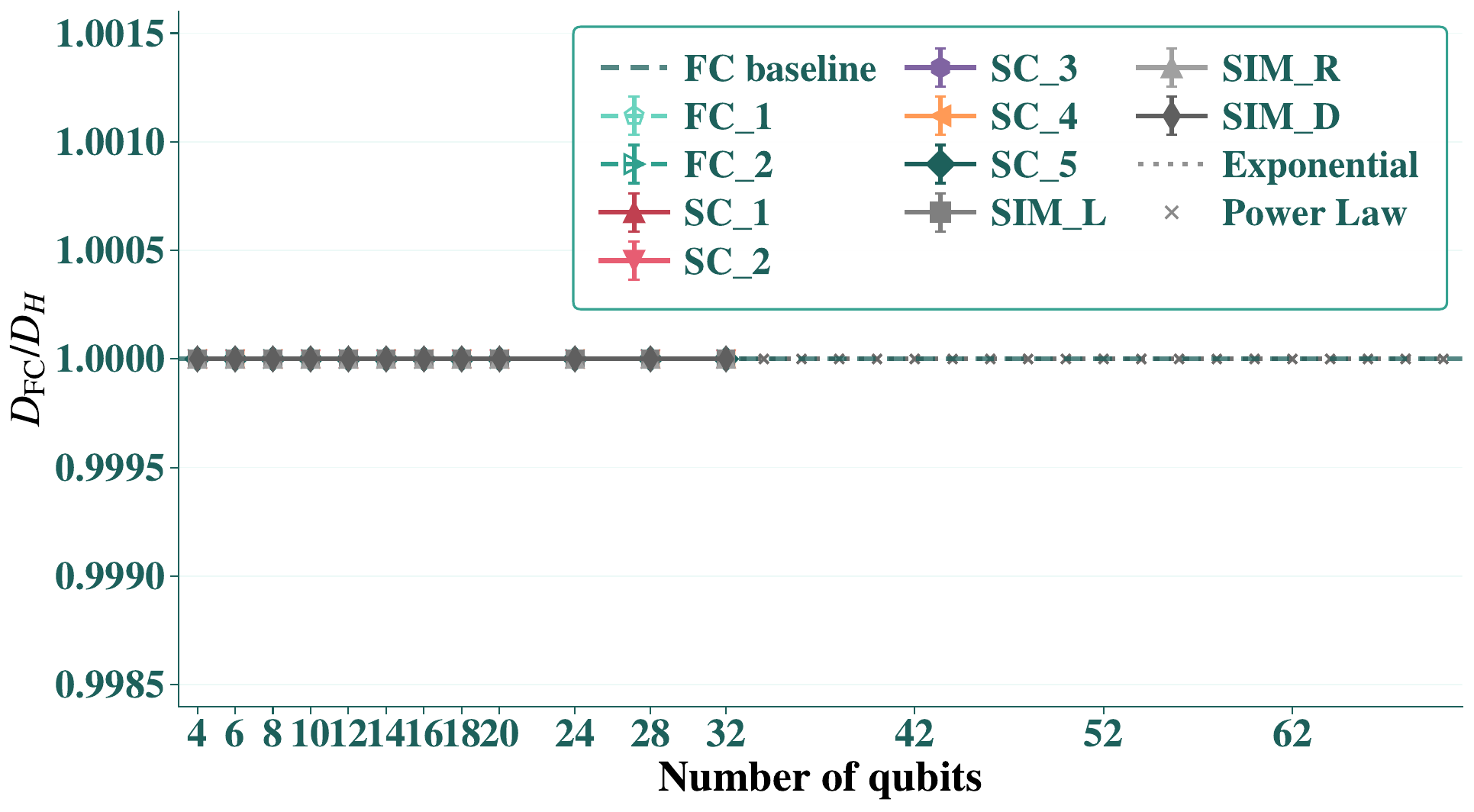}
        \caption{Local: nearest-neighbor only}
        \label{fig:ratio-local}
    \end{subfigure}
    \hfill
    \begin{subfigure}[b]{0.44\textwidth}
        \includegraphics[width=\textwidth]{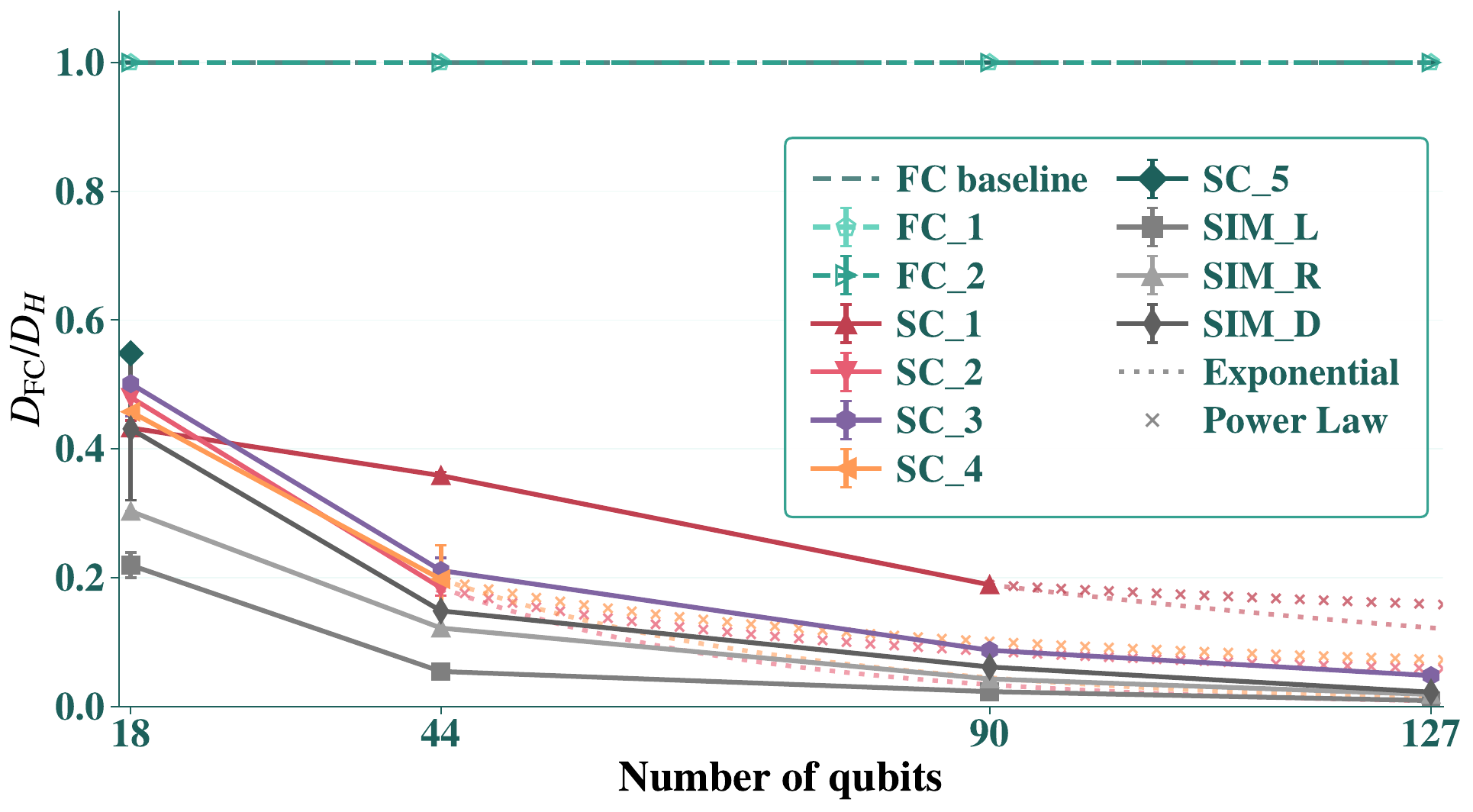}
        \caption{RHG lattice: $O(n)$ interactions ($n{=}18$--$127$)}
        \label{fig:ratio-rhg}
    \end{subfigure}
    
    \caption{Compilation efficiency $\eta_H=D_{\mathrm{FC}}/D_H$ versus qubit
count. Markers show means over ten circuit instances, with error bars
indicating one standard deviation; connecting segments show measured data.
Dotted lines and $\times$ markers show anchored exponential and power-law
extrapolations, respectively, fitted using up to the last ten measured sizes.
Both models are constrained to be non-increasing and remain positive.
Panels~(a)--(c) use measured data for $n=4$--$32$, while panel~(d) uses valid
RHG sizes $n\in\{18,44,90,127\}$.}

    \label{fig:compilation-overhead-scaling}
\end{figure}

\begin{figure}[t]
    \centering

    \begin{subfigure}[t]{0.31\columnwidth}
        \centering
        \includegraphics[width=\linewidth]{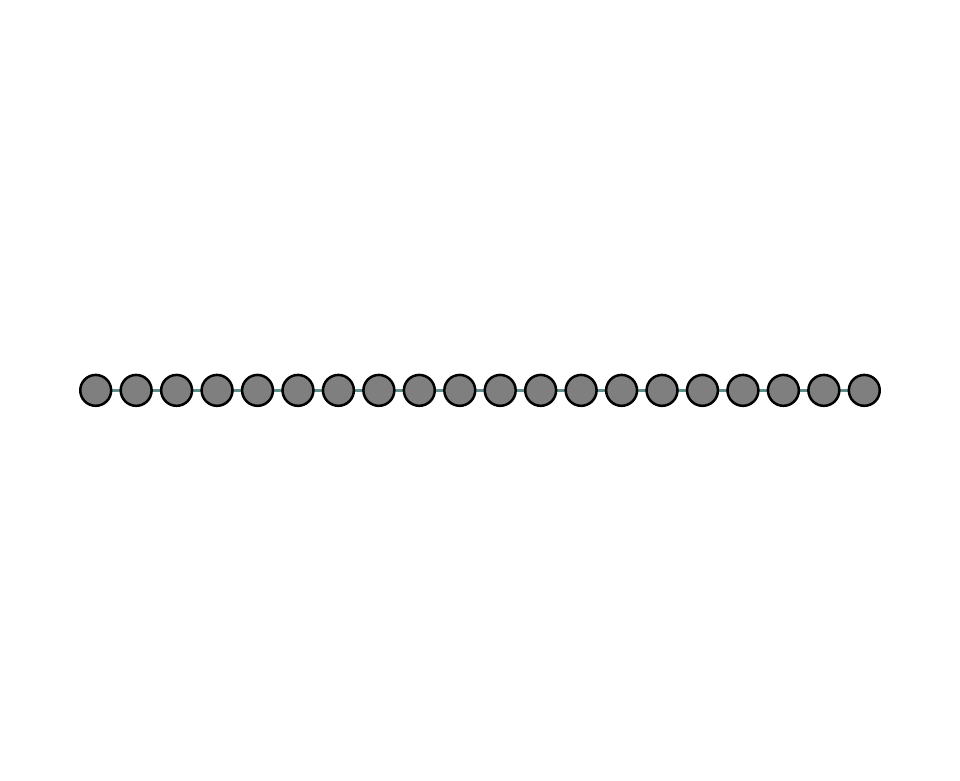}
        \caption{$SIM_L$: 1D nearest-neighbor chain.}
    \end{subfigure}\hfill
    \begin{subfigure}[t]{0.31\columnwidth}
        \centering
        \includegraphics[width=\linewidth]{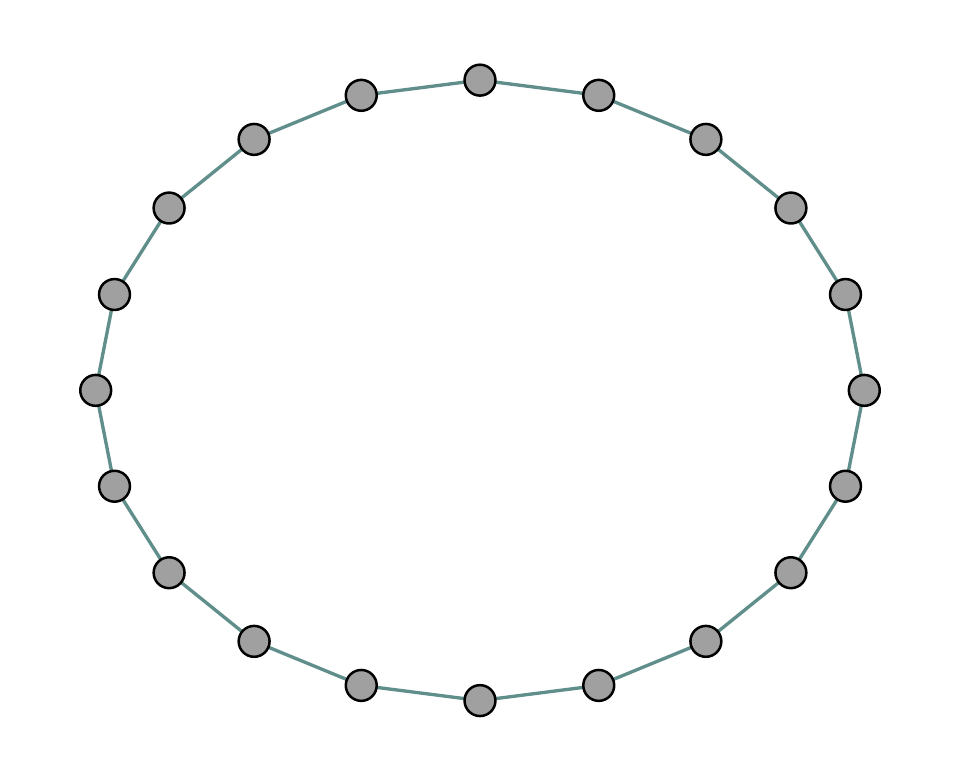}
        \caption{$SIM_R$: Ring, periodic 1D chain.}
    \end{subfigure}\hfill
    \begin{subfigure}[t]{0.31\columnwidth}
        \centering
        \includegraphics[width=\linewidth]{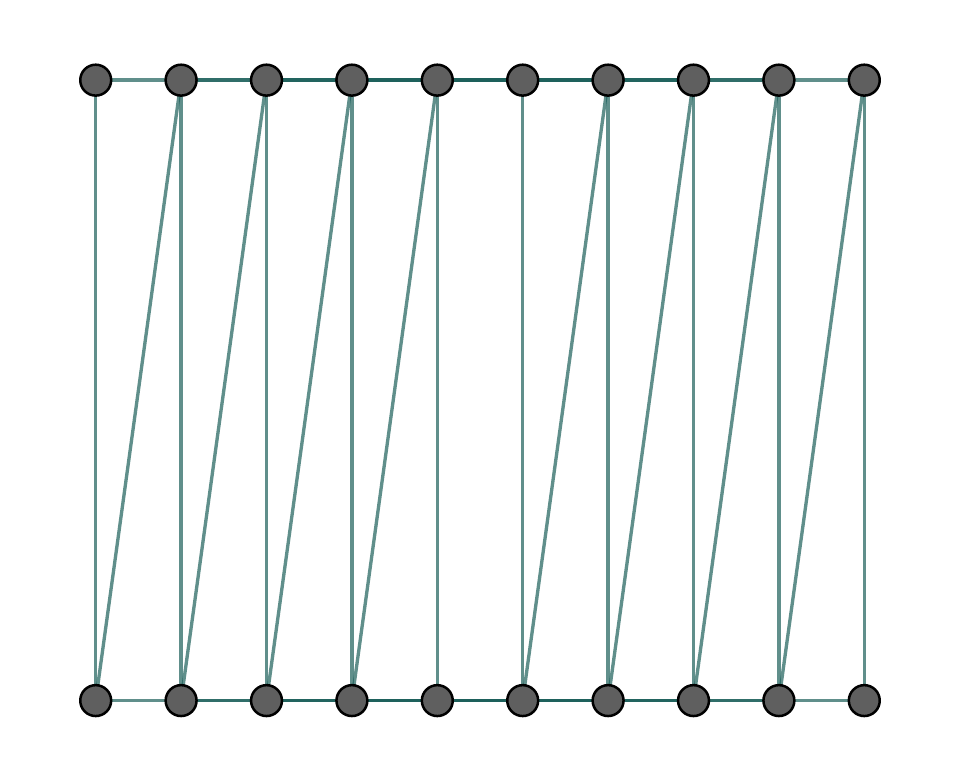}
        \caption{$SIM_D$: ladder, quasi-1D two-rail lattice.}
    \end{subfigure}

    \caption{Reference connectivity graphs used as simplified baselines. These idealized low-dimensional topologies provide controlled comparisons against the native hardware layouts in Fig.~\ref{fig:hardware_topologies}.}
    \label{fig:reference_topologies}
\end{figure}
\subsubsection{Setup}
For the dense, local, and sparse patterns, we compile IQP circuits for $n \in \{4,6,8,10,12,14,16,18,20,24,28,32\}$ across eleven connectivity graphs: the seven hardware targets of Fig.~\ref{fig:hardware_topologies}, plus fully connected, line, ring, and ladder reference topologies (Fig.~\ref{fig:reference_topologies}). We generate ten random instances per configuration for all $n$. For the RHG pattern, we compile the valid lattice sizes $n \in \{18,44,90,127\}$ with ten random instances per size. The plotted efficiency ratio is depth-based,
\[
\eta_H := \frac{D_{\mathrm{FC}}}{D_H}.
\]

\subsubsection{Results}
Fig.~\ref{fig:compilation-overhead-scaling} shows $\eta_H$ versus $n$ for all
four interaction patterns\footnote{Let $n_0$ be the largest measured size.
Using up to the last ten sizes, unweighted least-squares fits of
$\log\eta_H$ against $n$ and $\log n$ give slopes $b_{\rm e}$ and $b_{\rm p}$.
For $n>n_0$, we use
$\widehat{\eta}_{\rm e}(n)=\eta_H(n_0)e^{\min(b_{\rm e},0)(n-n_0)}$ and
$\widehat{\eta}_{\rm p}(n)=\eta_H(n_0)(n/n_0)^{\min(b_{\rm p},0)}$.
Both remain positive and decrease toward zero for negative fitted slopes.}. For dense instances (Fig.~\ref{fig:ratio-dense}), the fully connected trapped-ion-like experimental models remain at $\eta_H=1$ across the entire measured range, as expected from their all-to-all connectivity. By contrast, locality-constrained architectures show a clear degradation with increasing $n$. At $n=16$, the dense efficiencies already range from $\eta_H \approx 0.27$ for SC\_2/SC\_5 to $\eta_H \approx 0.47$ for SC\_3, with SC\_1 and SC\_4 at approximately $0.35$ and $0.32$, respectively. By $n=32$, the separation is stronger still: SC\_2, SC\_1, SC\_3, SC\_4, and SC\_5 reach $\eta_H \approx 0.15$, $0.27$, $0.18$, $0.20$, and $0.14$, respectively.

For local patterns (Fig.~\ref{fig:ratio-local}), $\eta_H=1$ across all platforms and all measured qubit counts, consistent with the zero routing overhead observed in the phase-diagram analysis: nearest-neighbour interactions embed directly into every connectivity graph considered here. Sparse patterns (Fig.~\ref{fig:ratio-sparse}) interpolate between the dense and local extremes. At $n=16$, the constrained hardware efficiencies lie between roughly $0.29$ and $0.43$; by $n=32$ they fall further, to approximately $0.17$--$0.28$ across SC\_1, SC\_2, SC\_3, SC\_4, and SC\_5.

The RHG pattern (Fig.~\ref{fig:ratio-rhg}) extends the analysis to device-relevant scales. At $n=18$, the fully connected platforms again remain at $\eta_H=1$, while the constrained hardware targets lie in the range $\eta_H \approx 0.43$--$0.55$. At $n=44$, FC\_1 still remains at unity, while SC\_1, SC\_3, SC\_2, and SC\_4 drop to approximately $0.36$, $0.21$, $0.18$, and $0.20$, respectively. At $n=90$, FC\_1 ($\eta_H=1.0$), SC\_1 ($\eta_H \approx 0.189$), and SC\_3 ($\eta_H \approx 0.087$) remain among the plotted device-like models. At $n=127$, SC\_3 reaches $\eta_H \approx 0.048$. The RHG trend is therefore especially informative at larger scales: although the interaction count remains only $O(n)$, embedding a three-dimensional interaction graph into low-dimensional hardware connectivity induces a strong routing penalty on sparse architectures.

\subsubsection{Implications for Simulatability}
As $\eta_H$ decreases, compiled depth grows relative to the fully connected
baseline and tightens the effective noise budget through
Eq.~\eqref{eq:noise_budget_ratio}. For dense and sparse $n{=}32$ circuits,
routing moves SC\_2, SC\_3, and SC\_4 across the selected boundary, while
SC\_3 also crosses for the RHG instances. These results reinforce the need
to consider interaction structure, effective noise, and connectivity jointly.
\section{Discussion}
\label{sec:discussion}

Our results support the framework of Sec.~\ref{sec:ciss}: limited
connectivity increases compiled depth from $D_{\mathrm{FC}}$ to $D_H$,
reducing the effective noise budget compatible with a positive
simulatability margin. Since
$D_\ast(p,k)=O\!\left(p^{-1}\log(k/p)\right)$, routing-induced depth
inflation acts approximately as a multiplicative penalty on the tolerable
noise level, up to the logarithmic correction.

\subsection{Connectivity Penalty and Noise Budget}

A central outcome of the phase-diagram analysis is that connectivity matters by converting a given native fully connected depth $D_{\mathrm{FC}}$ into a larger hardware-constrained depth $D_H$. This depth inflation reduces the noise budget compatible with $m(H)>0$. For the local pattern, where routing overhead vanishes and
$D_H=D_{\mathrm{FC}}$, SC\_1, SC\_2, SC\_3, and SC\_4 remain in the ``Potentially hard'' region, while SC\_5 lies in the sufficient simulatable region. Thus, the local results isolate the effects of native
compilation depth and the assigned noise proxy before any connectivity-induced routing overhead is introduced.

For dense and sparse $n{=}32$ circuits, sparse-connectivity models incur
substantial depth inflation, and routing moves SC\_2, SC\_3, and SC\_4
across the selected boundary. SC\_3 also crosses for the RHG instances.
SC\_1 remains below but close to the boundary, while SC\_5 is already in the
sufficient-simulation region at its fully connected baseline. Connectivity
therefore does not alone determine the classification, but it can reduce the
margin and, in selected cases, change it relative to the sufficient
simulation condition.

\subsection{Gate Speed versus Computational Hardness}

Superconducting platforms can execute two-qubit gates much faster than trapped-ion platforms~\cite{chen2025efficient, schafer2018fast, bruzewicz2019trapped}, but gate speed is not the quantity relevant to the simulatability boundary considered here. The relevant distinction is whether the compiled noisy circuit lies in the regime $m(H)<0$, where efficient classical approximation is expected, or in the regime $m(H)>0$, where this sufficient simulation criterion does not certify efficient classical simulation.

The crossings in Fig.~\ref{fig:phase_diagrams} show that faster execution
does not by itself compensate for routing-induced depth inflation and finite
noise. Figure~\ref{fig:compilation-overhead-scaling} further indicates that
the margins of sparse-connectivity models decrease with system size. The
relevant resource is therefore the ability to keep compiled depth below the
noise-dependent critical scale, not gate speed alone.

\subsection{Scaling to \texorpdfstring{$\mathcal{O}(10^2)$}{O(10to2)} Qubits}

The depth-scaling data in Fig.~\ref{fig:compilation-overhead-scaling} show that the connectivity penalty becomes more severe as system size increases. 
Because $\eta_H=D_{\mathrm{FC}}/D_H$, any reduction in $\eta_H$ means that the hardware-constrained circuit consumes a larger fraction of the available depth budget than its fully connected counterpart. Equivalently, maintaining the same simulatability margin requires a correspondingly smaller effective noise rate according to Eq.~(\ref{eq:noise_budget_ratio}). Connectivity therefore acts as an approximately multiplicative penalty on the tolerable noise threshold, up to the logarithmic correction in the critical-depth scaling.

For the sparse-connectivity models considered here, this penalty becomes
substantial at device-relevant scales. Even though the RHG interaction graph
contains only $O(n)$ edges, its hardware embedding can dominate compiled
depth, tightening the noise reduction required as the system grows.

\subsection{Noise Model Limitations}

Our analysis uses $p_{\mathrm{eff}}(H)$ as a simplified effective-noise proxy based primarily on reported two-qubit gate error rates. This is a useful first-order choice, but the total noise affecting a compiled IQP circuit is not limited to two-qubit errors alone.

Connectivity-constrained compilation can also amplify memory errors through longer circuit duration and idle time, and can worsen crosstalk; these effects compound the usual SPAM and one-qubit contributions already present in the device error budget~\cite{cowtan2019qubit, das2021adapt, rudinger2021experimental, bravyi2021mitigating}. Accordingly, the connectivity-induced shift in simulatability should be interpreted as arising from overall noise exposure, with the present $p_{\mathrm{eff}}(H)$ serving as an approximate proxy. An important direction for future work is to extend this framework to broader error models and real quantum devices, and to integrate error mitigation and pre-fault-tolerant error-suppression techniques in order to study how they modify the estimation of $p_{\mathrm{eff}}(H)$ and the resulting simulatability margins~\cite{Cai_2023, placidi2026deeplearningapproachesquantum, Cirstoiu_2023}. While such methods may improve the effective operating point, they do not eliminate the routing-induced depth overhead associated with limited connectivity.

\section{Conclusion}
\label{sec:conclusion}

We have systematically analyzed the connectivity-induced simulatability shift for noisy IQP circuits, connecting the theoretical framework of~\cite{rajakumar2025polynomial} to empirical compilation data across seven quantum device models.

Under the adopted effective-noise proxy, routing moves SC\_2, SC\_3, and
SC\_4 across the selected boundary for the dense and sparse $n{=}32$
patterns, and SC\_3 for the RHG instances. No connectivity-induced shift
occurs for the local pattern. Fully connected platforms avoid routing
overhead, although their final margins still depend on native compilation
cost and effective noise.

These results support the central claim of this work: limited connectivity is a direct constraint on actionable quantum advantage in noisy IQP circuits. Sparse connectivity increases compiled depth, and because the critical boundary scales approximately as
$p^{-1}\log(k/p)$, this depth inflation forces a correspondingly lower hardware noise rate to preserve a favorable margin relative to the known simulatability boundary. In this sense, connectivity and noise must be treated jointly: reducing noise helps, but poor connectivity simultaneously increases the level of noise suppression required.

More broadly, the phase-diagram methodology developed here provides a practical way to connect hardware compilation data to noise-dependent simulatability criteria. Although this work focuses on IQP circuits, the same methodology can be applied to other circuit families whenever a noise-dependent classical simulatability boundary is available.\\

Code and processed data are available from the first author upon request.
\section*{Acknowledgment}
The authors thank Yi-Hsiang Chen, Marcello Benedetti, Harry Buhrman, and
Stephen Clark for helpful discussions.

\newpage
\bibliographystyle{IEEEtran}
\bibliography{refs.bib}

\end{document}